\documentclass[11pt]{article}
\usepackage{amsmath}
\usepackage{amsfonts}
\usepackage{graphicx}
\usepackage{hyperref}
\usepackage{url}
\usepackage{xcolor}
\usepackage{microtype}
\usepackage{booktabs}
\usepackage{fancyhdr}
\usepackage{lipsum}
\usepackage{siunitx}
\usepackage[T1]{fontenc}
\usepackage[a4paper, left=2.1cm, right=2.1cm]{geometry}

\title{Urban Sprawl Is Associated with Reduced Access \\and Increased Costs of Water and Sanitation}

\author{
  Rafael Prieto-Curiel\thanks{Complexity Science Hub, Metternichgasse 8, Vienna, Austria. \texttt{prieto-curiel@csh.ac.at}} \and
   Pavel Luengas-Sierra\thanks{World Bank, 1818 H Street, N.W., Washington, DC 20433, USA} \and Christian Borja-Vega\thanks{World Bank, 1818 H Street, N.W., Washington, DC 20433, USA}
}

\date{}

\begin{document}
\maketitle

\section*{Abstract}

Many cities are expanding in areas with scarce rainfall and limited water retention capacity, and are also becoming elongated and sprawled, making it harder to deliver services. This study quantifies the impact of urban form on access to water. We craft comparable urban forms for over 100 cities in Asia, Africa, and Latin America. For each city, we analyse the distance to the centre, one of the most critical features of cities. We introduce two metrics: remoteness, which quantifies the distance of any location to the city centre, and sparseness, a population-weighted average of all locations. We find that less remote areas have higher average income, are closer to critical infrastructure and have higher access to sewage and piped water. Sparser cities have higher water tariffs, lower proximity to critical infrastructure, and lower access to sewage and piped water. Finally, we model urban expansion under three scenarios: compact, persistent, and horizontal growth. When cities expand through compact growth rather than horizontal expansion, 220 million more people could gain access to piped water, and 190 million more to sewage services. 

\section*{Introduction}

{
Water scarcity is recognised as a significant, high-impact global challenge \cite{liu2017water}. Worldwide, nearly two billion people lack access to an improved water source that is accessible on premises, available when needed, and free from faecal and priority chemical contamination (safely managed) \cite{adams2020water}. The situation is particularly concerning in many low- and middle-income countries, where a large part of the population lacks access, and urban expansion is the fastest \cite{waterEthiopia}. Combined with inadequate sanitation and hygiene, it results in nearly one million deaths annually \cite{adams2020water}. 
}

{
Cities are locations where economies of scale make it easier to deliver services. In cities, existing infrastructure can be shared among a large population \cite{pumain2004scaling}. Based on human needs and service delivery, the design of cities successfully provides people with essential services and critical infrastructure, including access to water \cite{he2021future}. Today, 57\% of the world's population lives in cities, yet only 20\% of people who lack essential services reside in one \cite{world2021progress}. Despite their success in providing services, cities face many severe challenges. One out of every four cities experiences water scarcity, and many require financial assistance to ensure their inhabitants have access to water, particularly in Southeast Asia and Africa \cite{mcdonald2011urban}. Some may experience `day zero' (when the water supply is almost exhausted), which will affect poor households the most \cite{bischoff2020shape}. Additionally, extended dry periods and heavy rains are increasing in frequency and becoming harder to predict \cite{aboelnga2020assessing}. They alter the equilibrium between water demand and supply \cite{march2010suburbanization}, exacerbating water scarcity \cite{haddeland2014global}. 
}

{
However, the most pressing issue related to water scarcity is population growth \cite{BorjaVegaWaterSharedProsperity}. By 2050, seven in ten people are projected to live in a city, so the urban water demand is expected to double \cite{florke2018water}. In cities, nearly half of the population may face water scarcity by 2050 \cite{he2021future}, and one billion people may have access to less than 100 litres of sustainable water per person per day \cite{mcdonald2011urban}. Due to the population growth, cities will expand considerably. However, urban expansion is often unplanned, expanding across low-density residential areas and in remote locations with steeper slopes, resulting in elongated and sprawling areas \cite{gomez2019spatiotemporal}. The urban form (meaning the layout and structure of cities, including the arrangement of buildings, streets and open spaces, patterns of land use and density that shape how a city functions) is the hallmark of efficient planning \cite{prieto2023scaling}. Urban expansion has lasting implications for green spaces, flood resilience, and biodiversity conservation within urban ecosystems \cite{angel2020shape}. It decreases the capacity to manage water sustainably and increases the risk of flooding \cite{anderson2023city, balaian2024urban}. A compact and round urban form facilitates the provision of essential services, leading to the efficient use of resources, such as electricity \cite{ki2024impact}. Combined with population growth, many of the expanding cities are located in areas subject to extreme weather conditions, characterised by scarce precipitation and water stress (Figure \ref{AridityRain}). For cities expanding in these locations, water scarcity is a permanent challenge (details in the Supplementary Material, SM).

\begin{figure}[ht!] \centering
\includegraphics[width = 0.95\linewidth]{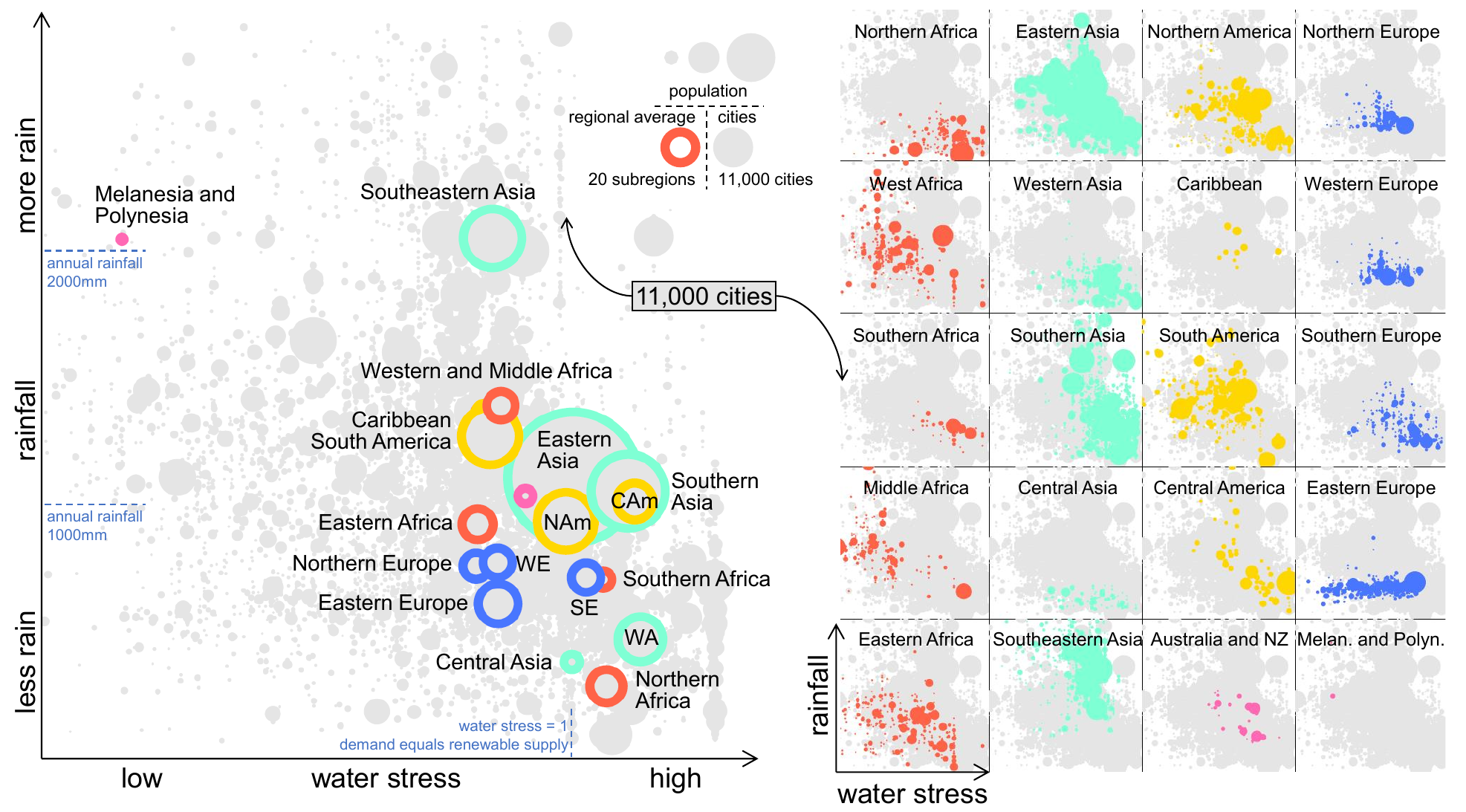}
\caption{Baseline water stress (horizontal axis) and yearly precipitation (vertical axis). The baseline water stress (from \cite{wri_aqueduct_2023}) measures the ratio of total water demand (including domestic, industrial, irrigation, and livestock uses) to available renewable surface and groundwater supplies. When the baseline water stress is equal to one, the water demand equals the renewable supply. Higher values indicate that withdrawals are larger than the available renewable water. The rainfall is the average monthly precipitation between 1970 and 2000, aggregated to a yearly value \cite{fick2017worldclim}. Most urban water is piped from reservoirs and water basins rather than sourced directly from rainfall, so values serve only as proxies to illustrate the vastly different contexts cities face in terms of available drinking water supply. Right - We compute both values for 11,000 cities (each represented by a different disc). The colour corresponds to the region, and the size of the disc corresponds to the population. Left - The rainfall and baseline water stress of each region are the average of cities weighted by the population. There are cities, mainly in Africa, Asia, and parts of America, where the baseline water stress is high and the rain is scarce. In those cities, granting access to water is more challenging. An interactive version is available in https://vis.csh.ac.at/urban-thirst/} \label{AridityRain}
\end{figure}
}

{
We quantify the effect of urban form on a wide array of local and city-level outcomes related to water and sanitation. New granular data has opened novel ways of quantifying the impact of the urban form \cite{depersin2018global, Tusting2019}. These data include terrain metrics, building stocks, satellite imagery, and other fine-grained information \cite{sirko2021continental}. Using detailed information on the footprint of buildings, we create comparable demarcations of more than 100 cities in Asia, Africa, and Latin America. For each urban area, we focus on the distance to the city centre, one of the most critical features of cities that relates to accessibility and other social patterns \cite{muth1961spatial}. To account for variations in population and urban form, we introduce the \textit{remoteness}, a comparable metric for the distance of any location within a city to its centre. We find that central areas are closer to built-up areas and to systems that provide essential societal services, such as transportation, energy, telecommunications, water, waste, education, and health (collectively known as critical infrastructure \cite{nirandjan2022spatially}). Remoteness also has a strong relation with access to services: when remoteness increases, access decreases substantially. In the most remote areas of a city, access to piped water is significantly lower than in the city centre. At the city level, the \textit{sparseness} is a population-weighted average of all locations within a city. We find that sparser cities have higher water tariffs, lower proximity to critical infrastructure, and lower access to piped water and sewage. On average, in a city with twice the sparseness, water tariffs are 75\% higher, proximity to critical infrastructure is 40\% lower, and access to piped water is also 50\% lower.
}

{
The manner and location in which cities expand can either increase or reduce access to water and sanitation services. For our sample of cities, we model urban expansion under three hypothetical scenarios: compact, persistent, and horizontal growth, considering that the only element that changes when the population increases is the location of the expansion within the city. Results quantify the extent to which compact development would increase the share of the population with access to piped water and sewage. If cities expand through compact growth rather than horizontal expansion, 220 million more people could gain access to piped water, and 190 million more people could gain access to sewage services.
}

\section*{Results}

{
We focus on over 100 major cities in Asia, Africa, and Latin America, selected with the objective of representing the most urban populations from the regions and capturing a wide range of geophysical and economic conditions (details in the SM). For each city, we first delineate its perimeter using detailed information based on the location and footprints of buildings from the Google Open Buildings dataset \cite{sirko2021continental}. This definition enables us to apply the same criteria across cities using the proximity of infrastructure (as it is done elsewhere \cite{Africapolis}). In total, there are 650 million inhabitants and 183 million buildings in the selected areas. Then, we manually define its centre, which is frequently the main square or transport station (details are provided in the Methods section). Although there may be many candidates for the centre of the city, the objective is to identify remote areas, for which the precise location of the city centre is less critical \cite{lemoy2021radial}. The proportion occupied by buildings decreases for regions far from the centre, but there are vast heterogeneities between what is constructed and how that decreases with distance (Figure \ref{RadialFigure}). In São Paulo and Jakarta, for example, up to 50\% of the surface near the city centre is occupied by buildings, whereas in Cairo (with a similar population), less than 20\% of the surface near the centre is occupied by buildings. In general, African cities have a smaller constructed surface area near the centre, which increases urban sprawl.

\begin{figure}[h!] \centering
\includegraphics[width = 0.8\linewidth]{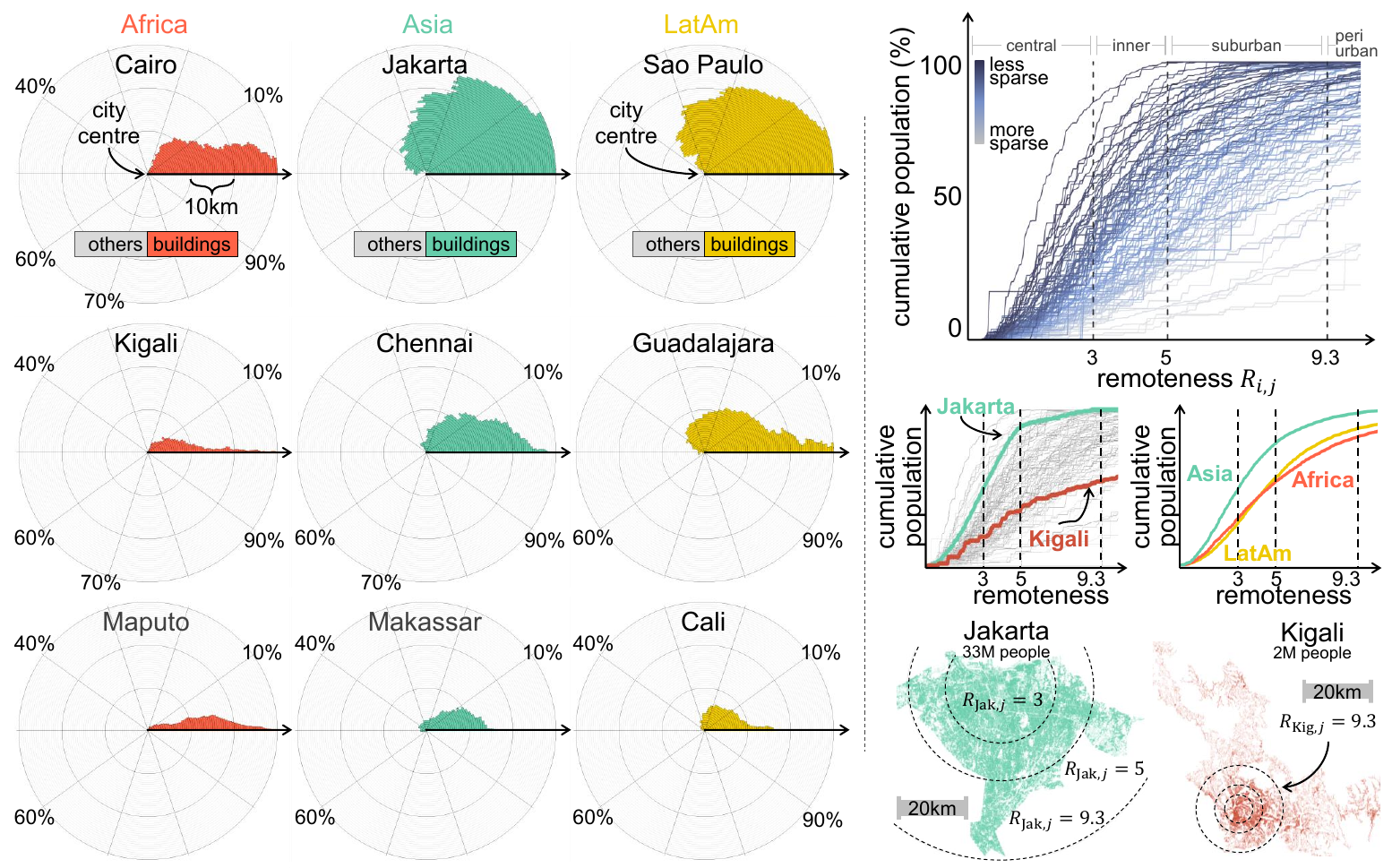}
\caption{Left - Fraction of the city's surface that is occupied by constructed buildings for some large (top), medium (centre) and small (bottom) selected cities in the study. Concentric circles represent each urban area, with the centre of the city aligned with the centre of the circles. Each additional ring corresponds to an extra 500 m from the centre. The coloured section of each ring represents the surface occupied by buildings. An interactive version for all cities in the study is available at https://vis.csh.ac.at/urban-thirst/. Right - Cumulative fraction of the population (vertical axis) that lives up to some value of remoteness (horizontal axis). Each line represents a city and its population, according to its level of remoteness. The figure below represents the cumulative population by remoteness for the three regions we studied. Also, the cumulative population for Jakarta and Kigali are highlighted, and the buildings are displayed on the same scale. In Jakarta, more than half of the population lives in the most central part of the city (regions with $R_{i,j} \leq 3$). However, in Kigali, only 15\% of the population lives in the central area.
} \label{RadialFigure}
\end{figure}
}

{
We use the centre of the city $i$ to characterise different locations \cite{lemoy2021radial}. Analysing locations within cities using concentric rings around the centre helps capture how cities expand, compact, or disperse in different directions from the core \cite{burgess1925growth}. This type of analysis can assess the impact of urban forms on housing, transport, electricity consumption, and other variables \cite{alonso1960theory, IUNGMAN2024e489}. However, the distance is not comparable across cities since it fails to account for urban sprawl (details in the Methods section). We construct a measure of \emph{remoteness} for location $j$ given by	
\begin{equation} \label{Remoteness}
R_{i,j} = 1000 \frac{ D_{i,j}}{\sqrt{P_i}},
\end{equation}
where $D_{i,j}$ is the distance in km from location $j$ to the centre of city $i$, and $P_i$ is the population of city $i$. Remoteness is a comparable index (with no units) for the location between cities since it discounts the general effect of population (details in the SM). To differentiate areas by their degree of remoteness, we group them into distinct categories. The classification is based on the constructed surface within walking distance. A tree is used to categorise the constructed surface within walking distance across the urban area (details in the SM). The motivation is to conceptualise the city as comprising four concentric zones, each containing a roughly equal amount of built-up surface. Locations with $R_{i,j} \leq 3$ fall within the \textit{central} part of the city, where 27\% of the walking-distance area is built-up. Then locations with $R_{i,j} \in (3,5]$ are in the \emph{intermediate} part, where 19\% of the area within walking distance is constructed. Places with $R_{i,j} \in (5, 9.3]$ are in the \emph{distant} part (and only 12\% of the surface within walking distance is constructed), and finally, with $R_{i,j} > 9.3$ are in the \emph{peri-urban} part, where only 7\% of the surface is constructed. The values of $R_{i,j}=3$, $5$ and $9.3$ were obtained by the classification process.  
}

{
The remoteness allows us to characterise and compare cities. For example, some cities tend to be highly sprawled, so most of their population is far from the city centre (Figure \ref{RadialFigure}). In Asian cities, more than half of the population lives in the central parts of the city, but in Latin America, only 28\%, and in Africa, less than 20\% of the population do. Notably, African cities are highly sprawled. Taking two buildings at random in Jakarta (with 33 million inhabitants), the average distance between them is 24.5 km, but taking two buildings at random in Kigali (a city with 2.2 million people), the average distance between them is 26.1 km. Kigali is much more sprawled, as the average distance is similar between both cities, but Jakarta is home to 15 times more people \cite{prieto2023scaling}. 
}

{
\paragraph{Remoteness and water.} We use gridded estimates of population and GDP per capita (representing income levels estimated at the municipality level) \cite{kummu2025downscaled}, as well as the Relative Wealth Index \cite{chi2022microestimates}, proximity to critical infrastructure \cite{nirandjan2022spatially}, and data for access to piped water \cite{JMPDatabase} to quantify the impact of different urban forms into various aspects on water and sanitation accessibility (details in the SM). Remoteness has a strong correlation with income, built-up area, and proximity to critical infrastructure within walking distance (Figure \ref{RemotenessByContinent}). When remoteness to the city centre decreases, the average income and proximity to critical infrastructure substantially increase. The average income of a person, proxied by the GDP per person in the municipality, is 30\% larger in the central part of the city than in the intermediate part, 56\% bigger than in the distant part, and nearly double than in the peri-urban part. Proximity to critical infrastructure is 30\% larger in the central part of the city than in the intermediate part, double than the distant part, and more than triple than the peri-urban part. Remote areas tend to be poorer and have limited access to urban amenities \cite{kummu2025downscaled}. The negative relationship between remoteness and income, as well as proximity to critical infrastructure, may be exacerbated in smaller cities (details in the SM). 
}

{
We aim to characterise the local conditions of each place, so we consider the built infrastructure within walking distance. We use 500 m since it is a threshold often used to approximate access to services for pedestrians \cite{SUGIYAMA2019100621}. However, we also take different thresholds, and the results are similar when the walking distance threshold varies (details in the SM). When remoteness to the city centre decreases, building infrastructure within walking distance also increases. For example, relative to the peri-urban part of the city, the constructed surface area is more than quadruple in the centre, the number of buildings is more than triple, and the building size is 19\% larger. Building infrastructure indicates whether an area has a high income or access to critical infrastructure. 
}

{
Remoteness also affects proximity to critical infrastructure. For pixel $k$, we fit the expression $I_{k} \propto R^{\alpha}_{i,k}$, where $I_{k} \in (0,1)$ is the infrastructure index of $k$ (with higher values indicating more infrastructure). We find that $\alpha = -0.682 \pm 0.008$, meaning that more remote locations have less infrastructure. At the local level, remote locations are often far from critical infrastructure. Keeping everything else constant, a pixel that is twice as remote as another has $2^{( - 0.682)} = 0.62$ times the critical infrastructure. 
}

{
Across continents, areas closer to the city centre are wealthier, closer to critical infrastructure, and have higher constructed surface area (Figure \ref{RemotenessByContinent}). Remote and isolated areas tend to be poorer and devoid of buildings and critical infrastructure.

\begin{figure}[h!] \centering
\includegraphics[width = 0.8\linewidth]{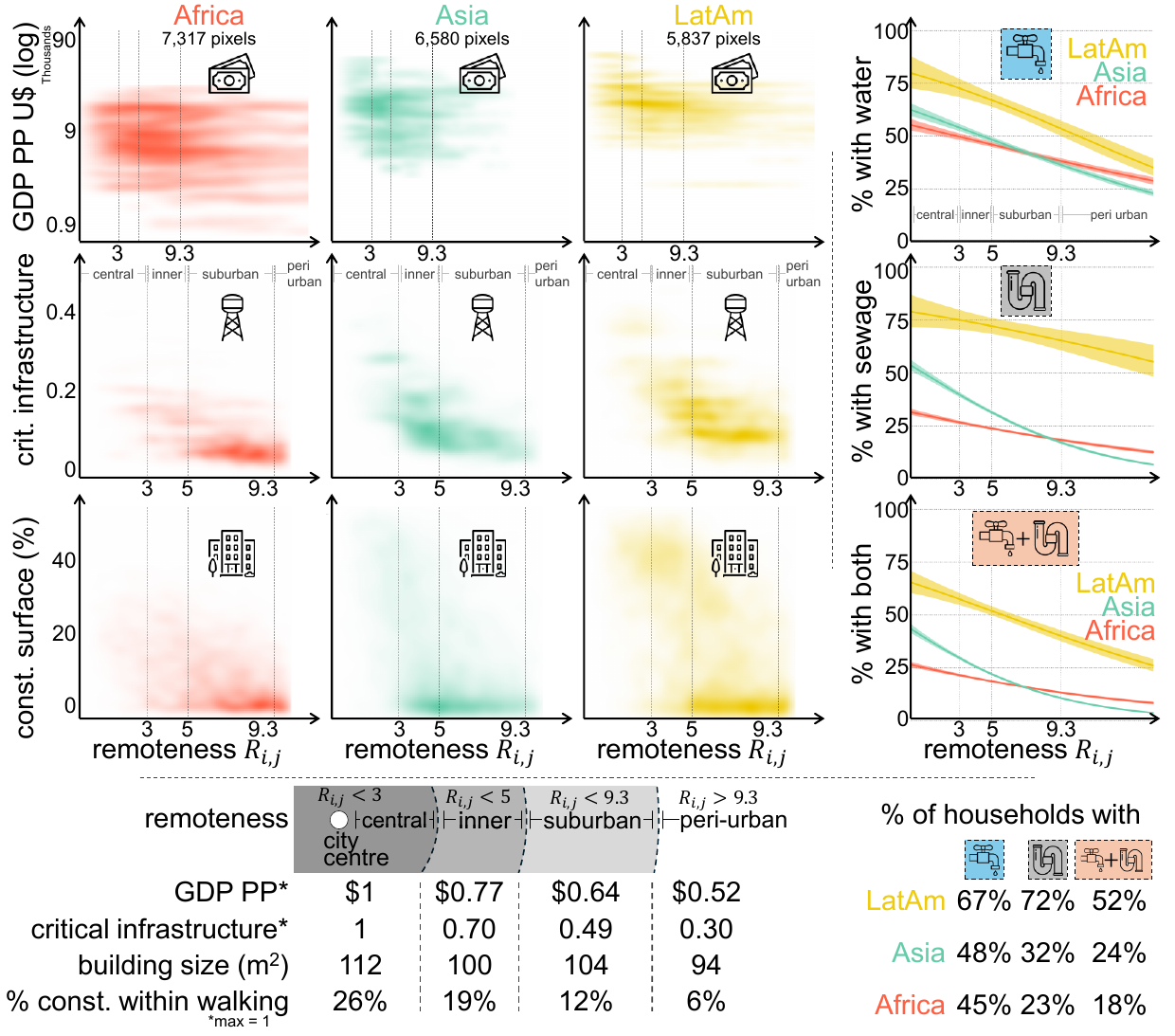}
\caption{Left - For 7,317 pixels in Africa, 6,580 in Asia and 5,837 in Latin America, we measure their remoteness (horizontal axis) and compare it to the income (vertical axis) \cite{kummu2025downscaled}, proximity to critical infrastructure \cite{nirandjan2022spatially}, and the constructed surface within walking distance \cite{sirko2021continental}. Right - Estimated frequency (vertical axis) in which households have access to water, sewage or both services, depending on the remoteness of each household (horizontal axis). Each line corresponds to the predicted probability obtained with a logistic model, taking remoteness as the independent variable. The colour indicates whether the model is for Latin America and the Caribbean (yellow), Asia (green), or Africa (red). A house in the centre of an Asian city, for example, has a 0.7 probability of having access to piped water, but it drops to less than 0.5 for a house in the suburban or peri-urban areas. The same decreasing trend is observed across all regions. The models were estimated using the Demographic and Health Surveys Program \cite{DHSData}. Bottom - GDP per person (based on purchasing power parity) and Critical Infrastructure compared against the centre of the city. Average building size in m\textsuperscript{2} and constructed surface (in \%) within walking distance for different areas of a city. } \label{RemotenessByContinent}
\end{figure}
}

{
In the development of a city, basic water and sanitation services may be unevenly distributed, with areas far from the urban core facing higher infrastructure costs \cite{coury2025value}. This process delays the provision of services, reduces land value and increases urban development costs, contributing to long-term disparities \cite{coury2025value}. Access to piped water, sewage, or both services decreases when remoteness increases. We utilise data from more than 125,000 household surveys collected across over 40 countries and 74 cities, as part of the Demographic and Health Surveys Program (DHS) \cite{DHSData}. For each household, the collected information indicates whether it has access to water and sanitation services and the location (in the form of GPS coordinates) of its neighbourhood. Using the remoteness of each household, we estimate the probability of having access to water and sanitation services (Figure \ref{RemotenessByContinent}). Across all cities, central areas are more likely to have access to water and sewage. For example, in Latin America, approximately 78\% of the households in the city centre have access to piped water, but only 65\% of the distant households and 53\% of the peri-urban households do. In Asia, 61\% of the households in the centre have access to sewage, but only 36\% of the distant households and 19\% of the peri-urban households do. Across continents and in remote locations, access to piped water is limited. In remote urban areas of Africa and Asia, sewage is also rarely available.
}

{
The way cities expand can reduce access to water and sanitation services. We design a framework to consider the impact of different urban growth scenarios on access to water and sewage. We analyse three hypothetical scenarios for urban growth representing contrasting trajectories: compact, persistent, and horizontal (details in the Methods section). In the three scenarios, we assume that the relationship between remoteness and access to services remains the same and consider that the population grows, increasing also the constructed surface (with no vertical expansion). What differs between the three scenarios is the location of the new constructions. In the compact scenario, the constructed surface grows radially from the centre through urban infill until each consecutive ring reaches a maximum potential expansion. In the persistent growth scenario, cities expand radially and each ring, at most, doubles its constructed surface. In the horizontal expansion scenario, we take the average constructed surface of the inner part and then let the city grow radially with that constructed surface ratio for each ring. A compact development, achieved through increasing density and urban infill, would increase the share of the population with access to piped water and sewage (Figure \ref{ThreeGrowthScenarios}). Persistent growth would result in a similar share of the population having access to water and sewage. However, horizontal and sprawling growth would decrease access to services.  

\begin{figure}[h!] \centering
\includegraphics[width = 0.65\linewidth]{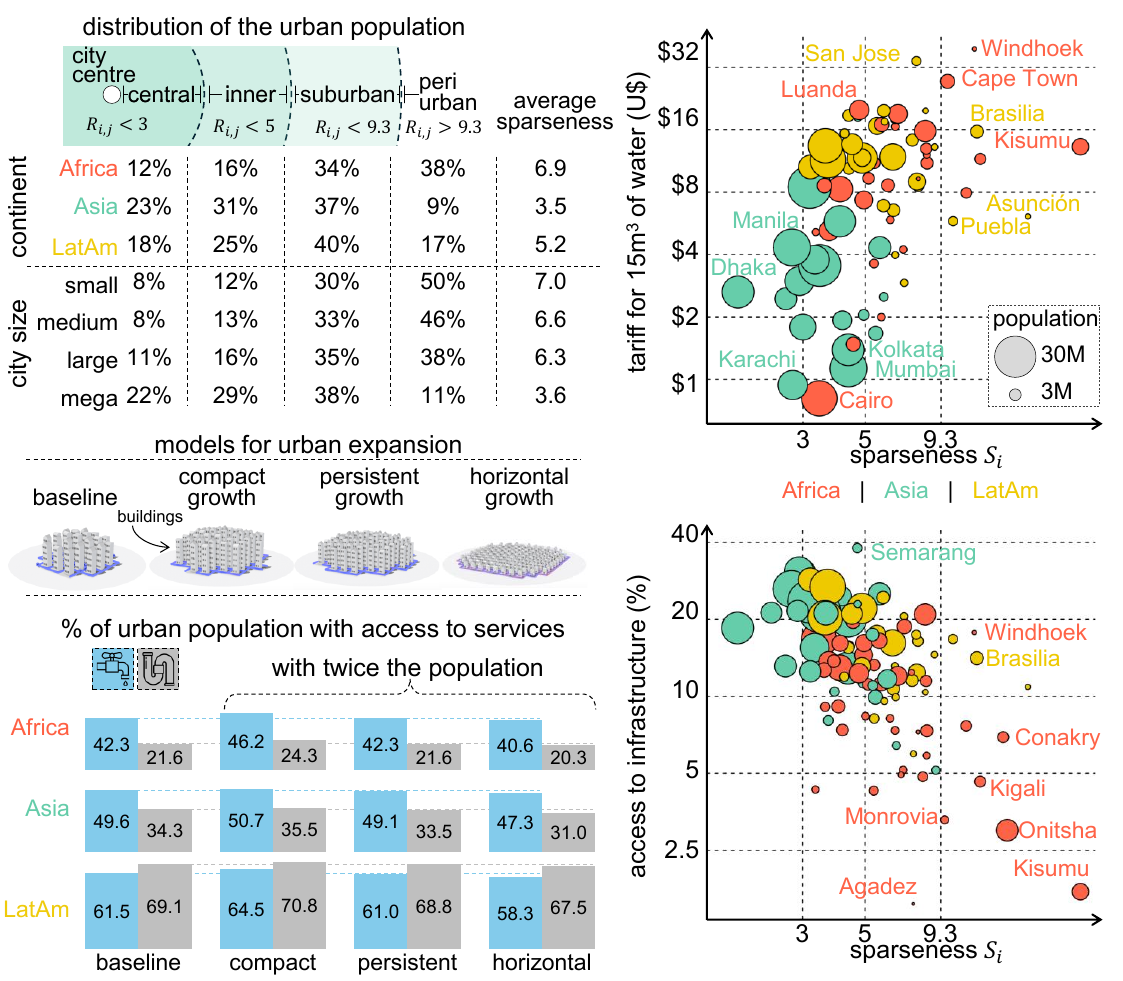}
\caption{Left - Distribution of the urban population across different parts of a city in Africa, Asia and Latin America. The average sparseness in Asia is only 3.5, whereas the average sparseness in Africa is 6.9, which is nearly twice as high. Three models for urban form expansion considering the impact of population growth: compact, persistent and horizontal expansion. To obtain the effect of urban expansion, we assume that the population and constructed surface of the city double, with the only difference being where the city's footprint expands, as indicated by the predicted probabilities in Figure \ref{RemotenessByContinent}. For the three scenarios, cities expand radially from the centre, as shown in Figure \ref{RadialFigure}. Estimated share of the urban population with access to piped water and sewage in Africa, Asia and Latin America, considering cities with twice the population. Three forms of urban expansion are analysed. One is where the urban expansion is from the city centre through infill development, one where urban growth follows the current population distribution, and one where expansion is mostly horizontal. Right - Sparseness (horizontal axis) and tariffs paid for 15 m\textsuperscript{3} of water \cite{WaterTariffs} and mean proximity to critical infrastructure (vertical axis). Water tariffs are influenced by a variety of factors, including scarcity, availability of infrastructure, and local economic conditions. Additionally, tariffs do not correspond with affordability, and there are significant disparities in affordability across and within cities.} \label{ThreeGrowthScenarios}
\end{figure}
}

{
By 2050, the urban population is projected to increase by 171\% in Africa, 53\% in Asia, and 30\% in Latin America \cite{UrbanizationProspects}. We use the average effect of urban expansion observed on each continent, combined with forecasted urban growth through 2050, to estimate future access to water and sewage services (see Methods for details). The principle is that by analysing cities across each continent, we can approximate the broader impact of urban expansion. For each continent, we apply the average observed effect of urban form to the projected urban growth to estimate how many people will gain or lose access to water and sewage. For example, in Africa, the average effect observed across 45 cities suggests that access to sewage would rise from 21.6\% to 24.3\% under compact growth, but fall to 20.3\% with horizontal expansion (Figure \ref{ThreeGrowthScenarios}). Whether cities expand through compact growth could result in granting 220 million more people access to piped water and 190 million more people access to sewage. In Africa, cities are expected to experience the most rapid growth. Due to rural migration, population growth, and the emergence of new agglomerations, the urban population in Africa is expected to increase by more than three times, from 550 million people in 2018 to nearly 1.5 billion by 2050 \cite{UrbanizationProspects}. Whether African cities expand compactly could mean granting 140 million additional people access to piped water and 100 million people access to sewage. Similarly, the urban population in Asia is expected to grow by more than 50\% by 2050. The way Asian cities expand could also grant 63 million people access to piped water and 84 million people access to sewage. Finally, the urban population in Latin America is expected to grow by 30\% by 2050. The way Latin American cities expand could grant 63 million people access to piped water and 84 million people access to sewage. Results need to be taken with care since compact growth can also result in high-density slums with worse access to basic services, even in central parts of the city \cite{pierce2017basic}.
}

{
\paragraph{Urban form at the collective level.} Derived from remoteness, a measure of \emph{sparseness} allows us to characterise the city as a whole. For city $i$, sparseness $S_{i}$ is defined as:	
\begin{equation} \label{SparsenessEq}
S_i = \frac{\sum_{j} P_{i, j} R_{i,j}}{ \sum_{j} P_{i,j}},
\end{equation}
where each location $j$ weighted by its population $P_{i,j}$. Cities in Asia tend to be less sparse compared to cities in Latin America and Africa (Table \ref{ThreeGrowthScenarios}). On average, 12\% of people in African cities live in the central area, and 38\% live in the peri-urban area. In contrast, 23\% of the population of Asian cities live in the centre and only 9\% live in the peri-urban area.   
}

{
African cities are nearly twice as sparse as Asian cities. Also, smaller cities tend to be more sparse (Figure \ref{ThreeGrowthScenarios}). In small cities (fewer than 1.5 million inhabitants), more than half of the population lives in the peri-urban area. In contrast, in megacities (with a population above 10 million inhabitants), only 11\% live in the peri-urban areas. On average, people living in a city with twice the population have 14\% less sparseness (details in the SM) . 
}

{
Many factors influence the determination of water and sanitation tariffs. However, the distance from the service point to the water source and the utility are essential factors in determining the service cost, mainly if tariffs are based on a cost-recovery model \cite{afdb2010}. Water tariffs are essential to cover the cost of providing access, but when they are high, fewer people can afford the service \cite{abramovsky2020study}. Using city-level information on water tariffs, we examine the relationship between water tariffs and sparseness \cite{WaterTariffs}. In sparser cities, the cost people pay for one m$^3$ of water tends to be much higher (Figure \ref{ThreeGrowthScenarios}). On average, water tariffs are 75\% higher in a city with twice the sparseness (details in the SM). Additionally, the problem of affordability can be compounded. In some cities with deficient water supply infrastructure, families often have to pay high prices for bottled water or for the installation of filtration devices \cite{rodriguez2017household}. Access to water and sanitation services decreases with sparseness. For each city, we estimate the population with access to piped water using the DHS surveys. Results show that a city with twice the sparseness provides access to piped water to less than half of its population (details in the SM). 
}

\section*{Discussion}

{
Cities offer economies of scale that facilitate resource distribution, yet deficient urban planning can hinder access to essential services \cite{BorjaVegaWaterSharedProsperity}. While water-specific policies are vital for sustainable freshwater use \cite{ravenscroft2022seeing}, addressing poor urban planning is also critical \cite{BorjaVegaWaterSharedProsperity}. These policies can help neutralise water scarcity by promoting recycling and allocating resources more efficiently \cite{WBhighDry}. However, policies that address deficient urban planning can also help tackle this challenge \cite{BorjaVegaWaterSharedProsperity}. Urban expansion has a relevant impact on the population, where feedback mechanisms also shape the interplay between urban form and infrastructure. People in remote locations are usually poorer and have less access to water and sanitation services. Sprawled cities tend to pay higher tariffs for water and may provide water to fewer people. Fragmented urban expansion, particularly at the periphery, reflects systemic and path-dependent development patterns \cite{pierce2017basic}. 
}

{
Using building footprint data, we delineated over 100 cities across 55 countries, employing consistent criteria that enabled cross-country comparisons. We manually selected each city's centre, not to define it precisely, but to identify remote areas. We introduce the concept of remoteness to measure distance from the city centre in a way that accounts for city size. This metric reveals clear patterns: income, population density, built surface, and access to infrastructure all tend to decrease with distance from the centre. On average, income drops from one dollar in the centre to half in the inner city, a quarter in the suburbs, and one-eighth in peri-urban areas. Central areas often have better infrastructure and services, although exceptions, such as underserved slums, exist \cite{pierce2017basic}. In contrast, remote areas typically face lower incomes, reduced access to water and sanitation, and environmental challenges, such as aridity and limited water sources, that raise service costs \cite{mcdonald2011urban}. While poverty alone does not prevent good hygiene \cite{rusca2017bathing}, its combination with limited access to services in remote areas increases health risks, including diarrhoea and malnutrition \cite{adams2020water}.  
}

{
At the city level, we define sparseness as the population-weighted remoteness of all locations, capturing how population is distributed relative to the centre. Sparse cities face distinct challenges compared to more compact ones. We find that smaller cities tend to be sparser, highlighting the need for context-specific planning. Asian cities are generally less sparse than those in Latin America and Africa. Sprawled urban areas, often the result of poor planning, lack adequate access to services such as healthcare, education, housing, and transport \cite{HealthcareBangladesh}. In sparse cities, fewer people reside near critical infrastructure, resulting in lower access to water and sanitation. These cities also tend to have higher water tariffs, though affordability varies widely (for example, Cancún and Kathmandu have similar tariffs, but Mexico's per capita income is nearly nine times higher). Low affordability remains a significant issue in sparsely populated and remote areas. While water subsidies aim to address this, they often miss the poorest households, who either lack access to piped water or consume less, resulting in subsidies that disproportionately benefit wealthier users \cite{abramovsky2020study}.
}

{
Urbanisation and demographic growth mean some cities may multiply their current populations \cite{100MillionCities}, with major implications for residents' livelihoods. The availability of water varies greatly across cities and can be assessed using indicators such as the aridity index, water scarcity, water resource availability, and water deficit. The expansion of cities in water-scarce areas is a growing trend that intensifies existing challenges \cite{he2021future, rathore2024water}. Rapid growth, especially in African cities, often occurs without adequate planning or services \cite{WaterArushaTanzania}. In the context of fast population growth, the expansion can lead to dispersed and disorderly city growth \cite{guan2023evaluating}. Our analysis shows that effective planning can significantly enhance access to water and sanitation. We modelled three growth scenarios—compact, persistent, and horizontal—for a city doubling in size. Assuming constant conditions, compact growth could provide piped water to 220 million more people and sewage services to 190 million, simply due to better spatial planning. Round, dense urban forms are more sustainable and serviceable. Compact, walkable neighbourhoods and taller buildings reduce pressure on peri-urban ecosystems, helping to conserve green space and mitigate pollution \cite{prieto2023scaling, behnisch2022rapid}. Still, densification is not a cure-all. Some densely populated areas, such as Kibera in Nairobi or Makoko in Lagos, remain severely underserved despite their central locations \cite{pierce2017basic}.
}

{
Freshwater is a dwindling resource, and rising demand exacerbates scarcity. Beyond population growth, shifts toward water-intensive diets, particularly meat and dairy, could significantly increase per capita water use \cite{islam2019world}. Although daily drinking water needs are minimal, producing the food consumed each day requires over 2,000 litres, and this demand is growing (see SM). As global consumption of water-intensive goods expands, it increasingly clashes with declining freshwater availability. While some services, such as electricity and internet, can benefit from novel technologies, no comparable breakthroughs exist for water and sanitation. Delivering these services in remote areas remains inherently costly and dependent on effective planning.
}

\section*{Methods}

{
Understanding urban form and its relationship with water scarcity requires the use of a wide array of tools and indicators, including remote sensing, cellular automata models, and geographic information systems, coupled with a water evolution and planning model \cite{Hannon2022}. New granular data has opened novel ways of observing cities at a global scale \cite{depersin2018global, BigDataPolicyProblems, Tusting2019}. These granular data include building stocks, terrain metrics, road infrastructure, and natural resources \cite{OpenStreetMap, sirko2021continental, esch2022world, NASAJPL2020}. Suggested indicators to capture water scarcity typically include population, water use, and water availability \cite{liu2017water}. We model urban form using information on the footprint of buildings. The information allows us to generate the boundaries of cities. The resulting urban delineations are comparable across cities because they are defined using the same criteria. Then, we estimate the remoteness of each location within the urban area. For each city, we estimate its sparseness, which is a weighted mean of the remoteness of all its locations. As indicators, we use location-level indicators of income, built-up area, access to services, and proximity to critical infrastructure. We also use city-level indicators of water tariffs, access to services, and proximity to infrastructure. Using these tools and indicators, we model urban expansion and its effects on access to services. Described below are the tools and indicators.
}

\subsection{Modelling urban form: Demarcating each city using information of the footprint of buildings}

{
The analysis considers 105 cities. Major cities, state capitals, and other medium-sized cities were selected (details in the SM). Data limitations restricted the sample. For example, data on building footprints for China or Saudi Arabia was not available. Of the 105 selected cities, 45 are located in Africa, 30 in Latin America, and 30 in Asia.
}

{
Information on the footprint of buildings was extracted from the Google Open Buildings dataset (\href{https://sites.research.google/open-buildings/}{https://sites.research.google/open-buildings/}). The footprint of buildings dataset was obtained using a deep learning model with high-resolution satellite imagery (50 cm pixel size) \cite{sirko2021continental}. For each footprint, the dataset includes its centre (latitude and longitude), the area in m$^{2}$, and geometry (polygon). We process and analyse this information using R Statistical software \cite{RCoreTeam2018}. In total, we analyse the footprint of 183.4 million buildings, corresponding to a constructed surface nearly the size of Kuwait.
}

{
Demarcating urban forms is challenging \cite{arcaute2015constructing}. It can entail making many decisions on techniques and parameters \cite{rozenfeld2008laws, cottineau2017diverse, rozenblat2020extending}. We determine the boundary of each city using a consistent approach across all cities (Methods Figure \ref{HullCenter}). The approach relies on the building footprint and the proximity between buildings. Polygons of buildings join other polygons of buildings when they are less than 200 m apart. The same criteria have been applied elsewhere to construct the delineation of thousands of cities in Africa \cite{Africapolis}. To ensure consistency, we utilise a geometric structure known as the alpha-convex hull, a common tool in computational geometry. This structure depends on a parameter $\alpha$ that captures the degree of concavity of the boundaries. The process starts by defining points, which are the buildings in each city. Then, the standard convex hull of all points is computed. Constructing the alpha-convex hull entails modifying the standard convex hull by allowing the edges of the original shape of a concave region to be included or removed based on distance ($<200$ m) and the $\alpha$ parameter. 

\begin{figure}[h!] \centering
\includegraphics[width = 0.6\linewidth]{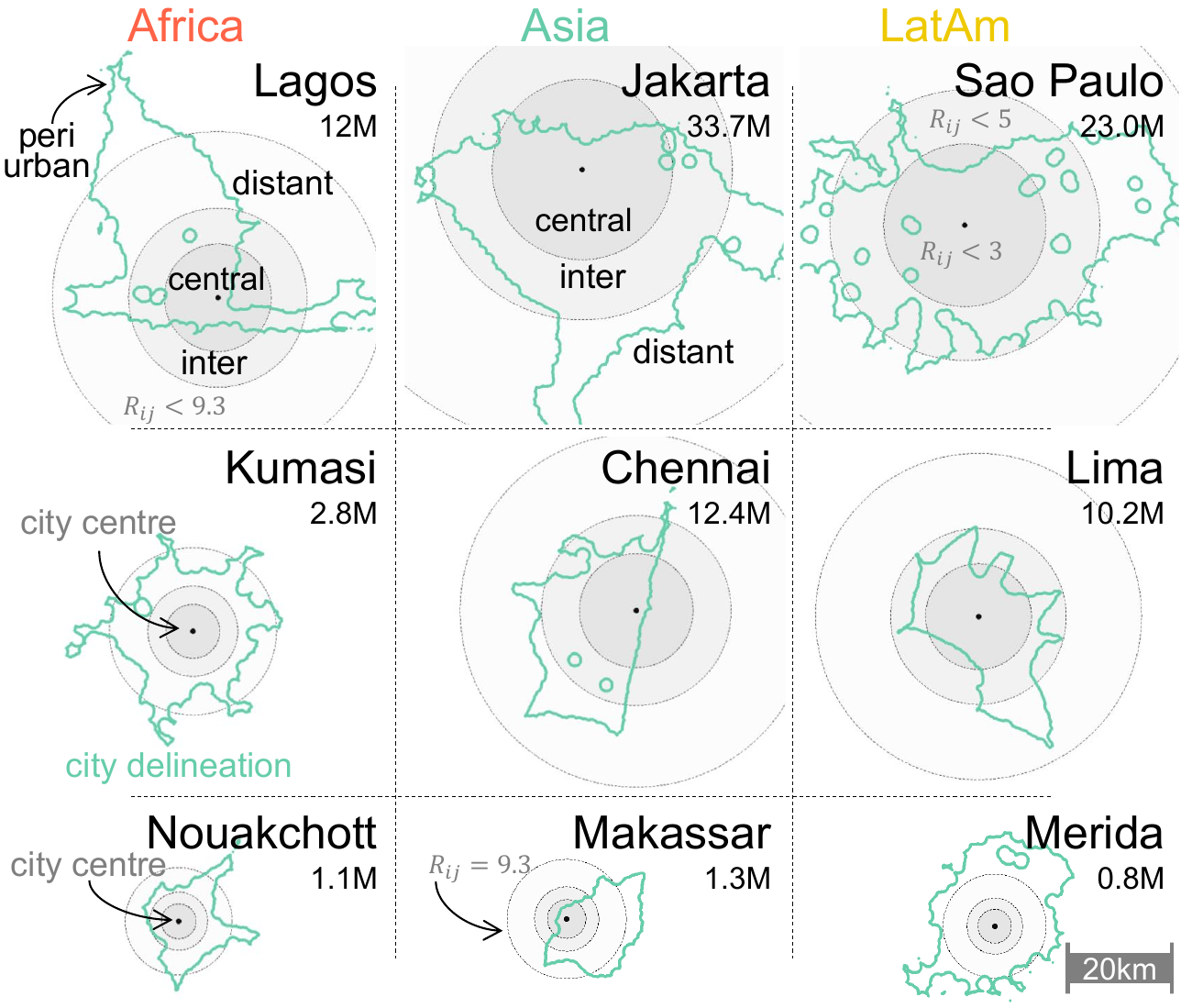}
\caption{The delineation of cities depends on the proximity of the building footprints. For each city, we use the footprint of all its buildings and compute the alpha-convex hull \cite{RCoreTeam2018}. The delineation for each city is the green polygon, which may contain holes and may result in discontinuous patches. For each urban agglomeration, we identify its centre as the main square, train station or city hall. Then, from the centre of city $i$, we measure the Euclidean distance to location $j$ in km and compute the remoteness $R_{i,j}$, which is an index of distance comparable across cities of different population size. For each city, we highlight the parts of the city with $R_{i,j}\leq 3$, which are the central area, with $3<R_{i,j}\leq 5$ are the internal part, with $5<R_{i,j}\leq 9.3$ are distant, and $R_{i,j}> 9.3$ which are peri-urban. } \label{HullCenter}
\end{figure}
}

{
For each city, we manually identify the centre as the location of the main square, church, mosque, or central transport station (Methods Figure \ref{HullCenter}). This technique is a common practice in urban studies, where the centre is defined using a combination of geographic, economic, and historical criteria \cite{lemoy2021radial}. Although the precise location of the centre can be ambiguous, the aim here is to detect remote areas, for which the accurate definition has a limited impact (details in the SM). Consider, for example, the city of Bogota. Its centre could be, among other locations, Plaza Bolívar, Museo del Oro (500 m north), or Parque Metropolitano (600 m west). The result of using different locations is that Soacha (a neighbourhood in the southeast part of the city) is 16.0 km, 16.4 km, or 15.5 km away from each of the three locations. Thus, Soacha is a remote location, and according to the centre, the distance varies slightly. Our analysis aims to detect remote areas, which turn out to be equally remote regardless of the point considered to be the centre (more details in the SM). 
}

{
The point selected as the city centre is used to determine the relative proximity of any other location within the urban area. We then use that location and observe the distance to the centre in concentric rings. Using a similar technique, by examining concentric rings around the centres of European cities, it was found that per-person CO2 emissions were significantly lower in compact and high-density cities compared to green and low-density cities \cite{IUNGMAN2024e489}.
}

{
For location $j$, remoteness is a function of the Euclidean distance, $D_{i,j}$ in km to the centre of city $i$ (Methods Figure \ref{HullCenter}). Distances from cities with different populations cannot be compared directly. For example, in Bogota (a metropolitan area with nearly 9 million inhabitants), a location 2.5 km away from the city centre is still a central location. However, in Manizales (a city with 434,000 inhabitants), an area within the same distance is peripheral. A common technique for making comparisons possible relies on the distribution of distances rather than their values \cite{IUNGMAN2024e489}. However, this technique ignores the sprawl of urban forms. Consider two cities with the same population but different urban forms. One city is compact and round; the other is highly sprawled. For both cities, one could consider locations in the first quintile of the distribution of distances to be close to the city centre. However, this technique is misleading. In the sprawled city, the first quintile will cover a much larger area, not because the city is large but because it is sprawled.
}

{
To make distances comparable between cities of different sizes, we rescale them using their population. In general, distances in a city grow according to the square root of its population \cite{lemoy2021radial, prieto2023scaling}. The principle stems from the way distances increase within a circle. Consider a circle with area $A$ and take any two points inside it. The expected distance between them is $(128/45\pi) \times \sqrt{A}$. In a circle, distances grow proportional to the square root of its area. Similarly, if the area of a city grows with its population, distances in that city should grow proportional to $\sqrt{P_i}$. Remoteness $R_{i,j}$ between the city centre $i$ and location $j$ is a function of distance and population as expressed in the equation to compute remoteness. For example, a location in Bogota 2.5 km away from its centre has $R_{i,j} = 0.9$ while a location in Manizales at the same distance has $R_{i,j} = 3.8$. Thus, that location in Manizales is much more remote.
}

{
For city $i$, we measure the sparseness $S_i$ as the weighted average of the remoteness of all locations $j$ within the urban form. The weight is the population at each location. Both remoteness and sparseness are related to other urban indicators, such as density, but they aim to capture more details regarding the distribution of the population and infrastructure within the urban polygon. Imagine, for example, a city with $P = 1$ million inhabitants (so that the expression for remoteness is the average distance to the centre) occupying a circular surface with radius $r = 10$ km. The city has an area of 314.16 km$^2$, resulting in a density of approximately 3,200 inhabitants per square kilometre. If the population is homogeneously distributed, then the average distance to the city centre is $2/3  \, r \approx 6.6$ km, so the sparseness of that city under a homogeneous distribution is $S_H = 6.6$. However, if the population lives mostly near the city centre, for example, in a high-density core with taller buildings forming a pyramid-like structure \cite{lall2021pancakes, batik2025best}, it is possible to arrange the population such that it has a small sprawl, $S_P < 3$. Similarly, if the population lives predominantly along the edge of the city (forming a kind of ring), the average distance to the centre would be approximately the radius $r$ \cite{batik2025best}. Thus, the sprawl would be, at most, $S_R = 10$. Therefore, three different layouts of a city, forming a ring, a flat homogeneous distribution, or a highly populated centre, can have the same density of 3,200 inhabitants per square kilometre, but the sparseness varies depending on the population arrangement, from $S_P < 3$ to $S_R = 10$.
}

\subsection{Pixels within a city}

{
Once the delineation of a city and its centre are identified, the next step is to define all locations within each urban form. For this purpose, we use the grid of the Relative Wealth Index database. This publicly available dataset provides local-level estimates of wealth for all low- and middle-income countries \cite{chi2022microestimates}. The estimates are provided at a 2.4 km resolution grid and constructed by merging different datasets, including DHS surveys. Wealth scores cannot be compared across countries. However, the scores serve as a systemic way of comparing different locations across cities \cite{sartirano2023strengths}. Pixels, each geolocated, comprise the grid. We take the grid and retain only the pixels within the demarcated urban forms. These pixels thus become the locations (Methods Figure \ref{Luanda}). Depending on the size of the city, more or fewer pixels are considered. For instance, more than 1,000 pixels are analysed in New Delhi, more than 500 in São Paulo, and more than 200 pixels in Cape Town. Overall, nearly 20,000 pixels are analysed (7,317 in Africa, 6,580 in Asia and 5,837 in Latin America). 

\begin{figure}[h!] \centering
\includegraphics[width = 0.6\linewidth]{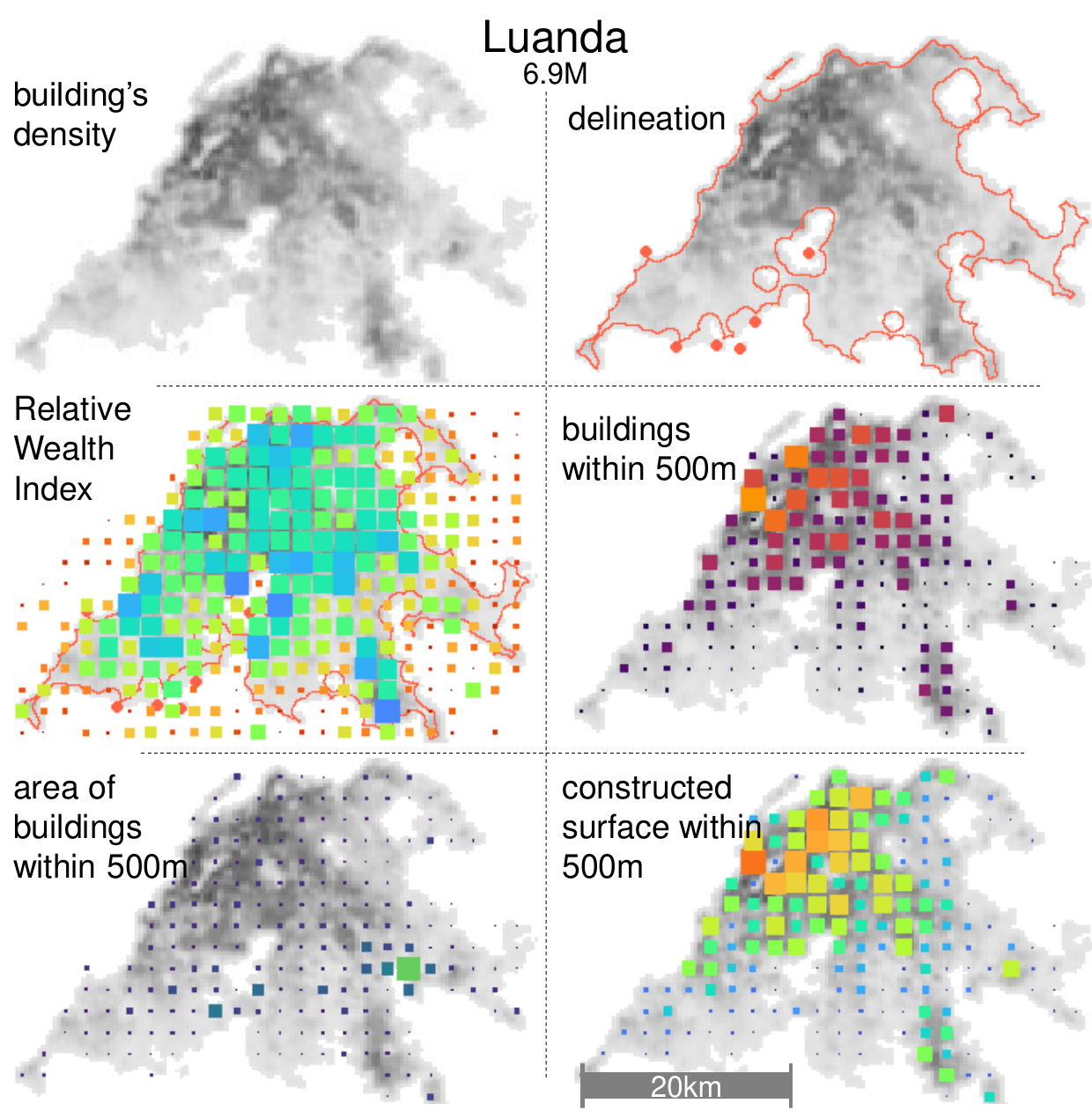}
\caption{The footprint of buildings (top, left) is used to construct the delineation of a city (top, right). Then, pixels within the city are considered separately based on the relative wealth index grid (middle, left). For each pixel, the number of buildings (middle, right), the average area of buildings (bottom, left) and the constructed surface within walking distance (bottom, right) are measured. The figure illustrates the case of Luanda, Angola, with a population of nearly 7 million inhabitants and a grid comprising 206 pixels.} \label{Luanda}
\end{figure}
}

\subsection{Location-level indicators} 
{
Besides remoteness, for each location (for each pixel), the following variables related to the footprint of buildings are estimated: constructed surface within walking distance, area of buildings, and number of buildings (Methods Figure \ref{Luanda}). Each variable consists of a sum within a radius deemed short enough to be within walking distance, here 500 m (see the SM for the sensitivity of results to varying the 500 m threshold). Also, for each location, we obtain its population from the gridded population dataset \cite{PopulationData, niva2023world} and its income from the gridded Gross Domestic Product dataset \cite{kummu2025downscaled}.
}

{
For each location, we analyse proximity to critical infrastructure using the Critical Infrastructure Index \cite{nirandjan2022spatially}. This dimensionless index ranges between 0 (no CI intensity) and 1 (highest CI intensity). It can serve as a proxy for future access to piped water and sewage. This is because neighbourhoods closer to critical infrastructure can be provided with access to piped water or sewage using fewer resources. Based on OpenStreetMap data, the index was constructed by examining 39 types of infrastructure systems, including transportation, telecommunications, waste, water, energy, education, and health. Because OpenStreetMap data depends on public contributions, it is incomplete and biased towards some locations \cite{prieto2022constructing}. However, the bias can be minor. Across all pixels ($n=19,734$), the critical infrastructure index shows strong correlations with aspects of the location, including GDP, remoteness, and constructed surface within walking distance (see the SM).
}

{
We correlate the remoteness with access to piped water and sewage. Information on local access to piped water and sewage is derived from 125,000 surveys conducted across 74 cities in Africa, Asia, and Latin America by the Demographic and Health Surveys Program \cite{DHSData}. First, using the information, we estimate household access to water, sewage, or both services. The percentage of households with water, sewage or both services varies considerably by continent.
}
 
{
Then, we rely on the location of the household to correlate its remoteness with access to services. Each household has information on the GPS location of its neighbourhood (cluster), allowing us to estimate its remoteness. Capturing a broad set of locations within each urban form, the sample at the city level is large (for example, the DHS has more than 15,000 respondents in New Delhi). For household $j$ in city $i$, we model the probability $\pi_{i,j}$ that it has access to piped water, to sewage, or both services with the equation:
\begin{equation} \label{LogitModel}
\pi_{i,j} = \frac{\exp^{\alpha + \beta R_{i,j}}}{1+\exp^{\alpha + \beta R_{i,j}}},
\end{equation}
where $R_{i,j}$ is the remoteness of household $j$ and where $\alpha$ and $\beta$ are two model parameters. If $\beta < 0$, a higher remoteness correlates with a lower probability of having access to services. For Asia, Africa, and Latin America, the parameters $\alpha$ and $\beta$ are estimated separately. Also estimated separately is the probability of having access to piped water, sewage, or both services. 
}

{
Higher remoteness correlates with much lower access to services (Methods Table \ref{TableWaterSewageBoth}). For example, in cities in Africa, distant households ($R_{i,j} > 5$) have less than a 50\% chance of having piped water, less than a 25\% chance of having sewage, and less than an 18\% chance of having both. A similar pattern of decay in access to services according to remoteness is observed in cities in Asia and Latin America (although baseline access in Latin America is generally higher). Although remote locations are less likely to have access to services, there are significant differences in the number of people who live in remote places across continents. In African cities, remote locations are both unlikely to have access to services and are often densely populated.
\begin{table}
\begin{center}
\begin{tabular}{l c c c}
 Water & Africa & Asia & Latin America \\
\hline
$\alpha$       & $0.210^{***}$  & $0.505^{***}$  & $1.372^{***}$  \\
               & $(0.016)$      & $(0.015)$      & $(0.027)$      \\
$\beta$        & $-0.075^{***}$ & $-0.115^{***}$ & $-0.133^{***}$ \\
               & $(0.002)$      & $(0.002)$      & $(0.004)$      \\
\hline
AIC            & $71657.656$    & $67314.594$    & $26183.788$    \\
\hline
Sewage & Africa & Asia & Latin America \\
\hline
$\alpha$       & $-0.778^{***}$ & $0.143^{***}$  & $1.321^{***}$  \\
               & $(0.018)$      & $(0.016)$      & $(0.027)$      \\
$\beta$        & $-0.077^{***}$ & $-0.185^{***}$ & $-0.074^{***}$ \\
               & $(0.003)$      & $(0.003)$      & $(0.004)$      \\
\hline
AIC            & $56382.156$    & $58662.651$    & $25029.004$    \\
\hline
Both & Africa & Asia & Latin America \\
\hline
$\alpha$       & $-1.019^{***}$ & $-0.265^{***}$ & $0.635^{***}$  \\
               & $(0.020)$      & $(0.017)$      & $(0.025)$      \\
$\beta$        & $-0.092^{***}$ & $-0.200^{***}$ & $-0.112^{***}$ \\
               & $(0.003)$      & $(0.003)$      & $(0.004)$      \\
\hline
AIC            & $49068.140$    & $50994.519$    & $28820.773$    \\
\hline
Num. obs.      & $52,897$        & $50,459$        & $21,349$        \\
\multicolumn{4}{l}{\scriptsize{$^{***}p<0.001$; $^{**}p<0.01$; $^{*}p<0.05$}}
\end{tabular}
\caption{Model for water (top), sewage (centre) and both (bottom) depending on the remoteness.}
\label{TableWaterSewageBoth}
\end{center}
\end{table}
}

{
There are also some relevant results at the pixel level. Remote locations tend to be poorer, have a lower Relative Wealth Index, and are, in general, far from critical infrastructure (details in the SM). The results are consistent with previous research, which shows that remoteness plays a detrimental role in access to urban amenities \cite{AmenitiesAggregation}.
}

\subsection{City-level indicators} 

{
At the city level, we construct indicators of building area, tariffs, and access to piped water. For each city, we measure the number of constructions, the constructed surface, and the mean building area. The infrastructure in some cities is primarily composed of small buildings. Many indicators show that at a collective level, sparse cities provide less access to services, are more expensive and reduce accessibility. On average, people living in a city with twice the sparseness have nearly 40\% less proximity to critical infrastructure (details in the SM). For example, in Khartoum and Agadez, almost two of every three buildings have a footprint smaller than 20 m$^2$. In Kuala Lumpur and Singapore, less than 15\% of buildings are as small (details in the SM).
}

{
We use the IBNet Tariffs database to link water tariffs and urban form \cite{WaterTariffs}. The database contains water and wastewater tariffs at the city level in a comparable US\$ format per city. Using the information, we correlate water tariffs and sparseness (see the SM). We find that in a city with twice the sparseness, water tariffs are 75\% higher.
}

{
For each city, we estimate the population with access to piped water. The estimate is obtained from country data from the Joint Monitoring Programme for Water Supply, Sanitation and Hygiene (JMP) \cite{JMPDatabase}. For each city, access to piped water is assumed to be the share of the population in urban areas of the country with access to this service.
}

{
We then analysed the data at the city level. Results show that sparser cities tend to have fewer people with access to water, less critical infrastructure, and pay higher water tariffs (details in the SM). Additionally, larger cities tend to be less sparse, but on the contrary, cities with more buildings tend to be sparser (details in the SM). There are also some differences between continents, with Asia having the least sparse cities. 
}

\subsection{Modelling urban expansion and its effects on access to services}
{
Over the next three decades, many urban areas worldwide are expected to expand considerably \cite{100MillionCities}. We analyse three hypothetical scenarios whereby a city could grow and estimate the effect of urban expansion. The hypothetical scenarios are used to detect the potential expansion of each urban form, and are constructed by taking a population and a constructed surface that doubles (e.g. there is no vertical expansion). Across scenarios, what differs is where the constructed surface expands. The motivation here is not to forecast how cities will grow, but to contrast trajectories of urban growth. For city $i$, a ring $k$ is formed by adding 500 m from the centre. A city has up to $n$ rings, each with a constructed surface $s_{i,k}$ and a potential expansion, $p_{i,k}$. We assume that the occupied surface can occupy, at most, 50\% of the surface, as observed in the centre of many cities. For the first scenario, compact development, we compute urban expansion for each ring as a constructed surface equalling 50\% of the surface. This scenario increases density and urban infill. For the second scenario, persistent growth, we consider a uniform expansion of the footprint. Here, the urban form is conserved during population expansion. For ring $k$, we take that the constructed surface doubles. However, if $s_{i,k} > p_{i,k}$, then in ring $k$ it is not possible to double the constructed surface. In that case, we increase the constructed surface of the subsequent rings, $k + 1$, $k + 2$ and so on. This scenario preserves aspects of the urban form delineation. For the third scenario, horizontal expansion, we consider the average constructed surface of the first $l$ rings, $d_i = (1/l) \sum^{l}_{k=1} s_{i,k}$ and then proceed sequentially from ring $1$ to ring $k$. For each ring, if the constructed surface is smaller than the mean constructed surface $d_i$, then we take the average $d_i$ as the constructed surface of ring $k$. We follow this process sequentially until the constructed surface has doubled in constructed surface. This scenario increases sprawl. 
}

{
To estimate the effect of urban expansion, we simulate a population that doubles and sample the location of each person in that city. With the location of each person, we compute remoteness. We combine the results from the water availability model (Equation \ref{LogitModel}) and the remoteness of each person to estimate the probability of having access to piped water, sewage, or both services. For each city, we obtain the average share of the population with access to piped or sewage under the three scenarios of urban expansion. This number is not a forecast of people with access to water or sewage, but is used only to estimate the impact of urban form and the location-level indicators (derived in Equation \ref{LogitModel}).
}

{
At the continental level, we calculate the average share of the population with access to services across cities. This yields the effect of urban form at the continental scale. The effect reflects both the typical direction of urban expansion (horizontal or vertical) and the likelihood that a person has access to water when living far from the city centre. We then combine this effect with forecasted urban growth from 2018 to 2050, based on the World Urbanization Prospects \cite{UrbanizationProspects}. To get the difference between 2018 and 2050, we multiply the result for each continent by its forecasted urban growth (which is 171\% for Africa, 53\% for Asia, and 30\% for Latin America). Finally, the difference across the three scenarios of urban expansion is obtained by comparing the total of each by continent. Compared to horizontal expansion (the third scenario), in compact development (the first scenario), 218.5 million more people gain access to piped water, and 194 million people gain access to sanitation services, including sewage.
}

\subsection{Data availability}

{
The data used for the analysis are structured into three separate tables, containing information at the city level (105 cities), at the pixel level (20,000 pixels), and at the survey level (with nearly 125,000 survey respondents). The aggregate data and results tables can be downloaded at

https://github.com/rafaelprietocuriel/WaterAndCities

Visualisations regarding all cities in the study can be found at 

https://vis.csh.ac.at/urban-thirst/
}

\clearpage

\section*{Supplementary materials}

\subsection*{Cities considered in the analysis}

{
To detect the impact of urban form on water, we first select the cities for the study. With our selection of cities, we aimed to capture a wide range of conditions, sizes, and geographies with a focus on developing countries. The base for our analysis is the buildings dataset and their footprints from the Google Open Buildings dataset \cite{sirko2021continental}. The dataset is only available for some countries in Asia, Africa, and Latin America, and is not available for others, such as China, Saudi Arabia, or Iran. As a result, some major cities could not be included. We aim to detect the impact of distance on water accessibility, so we mainly analysed large cities in which distance plays a more critical role.  

If available, we analysed the capitals of each country (14 in Latin America, 10 in Asia, and 30 in Africa). In some large countries, such as Brazil, Mexico, India, and Indonesia, more cities were included to detect any significant departures. Other cities were also included in the analysis to consider some spatial elements, such as Agadez in Niger, a significant uranium mining city in the Sahara Desert. 

In total, we consider 105 cities, of which 45 are located in Africa, 30 in Asia, and 30 in Latin America (Supplementary Figure \ref{Map}). More than 620 million people reside in the cities selected for the study, which is nearly the same population size as that of Latin America or Europe. 

\begin{figure*}[h!] \centering
\includegraphics[width = 0.75\linewidth]{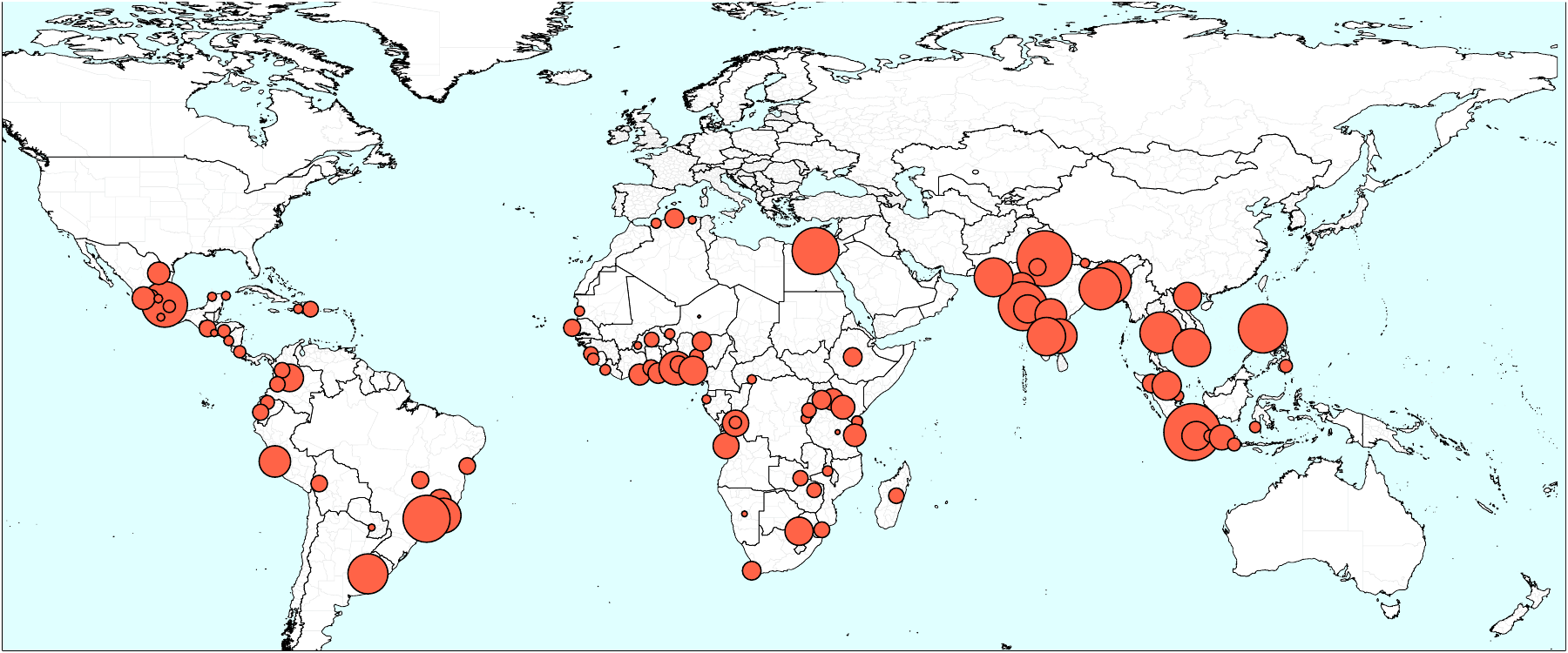}
\caption{Cities considered in the study. The size of the disc corresponds to the population of each city. The cities are \textbf{from Africa:} Algiers, Constantine, Oran, Luanda, Bobo-Dioulasso, Ouagadougou, Bujumbura, Bangui, Brazzaville, Kinshasa, Abidjan, Alexandria, Cairo, Addis Ababa, Libreville, Accra, Kumasi, Conakry, Kisumu, Mombasa, Nairobi, Monrovia, Antananarivo, Lilongwe, Nouakchott, Maputo, Windhoek, Agadez, Niamey, Aba, Abuja, Ibadan, Kano, Lagos, Onitsha, Kigali, Dakar, Freetown, Cape Town, Johannesburg, Dar es Salaam, Dodoma, Kampala, Lusaka, and Harare. \textbf{From Asia:} Dhaka, Phnom Penh, Ahmedabad, Bangalore, Chennai, Hyderabad, Jaipur, Kolkata, Mumbai, New Delhi, Pune, Surat, Bandar Lampung, Bandung, Batam, Denpasar, Jakarta, Makassar, Medan, Palembang, Semarang, Surabaya, Kuala Lumpur, Kathmandu, Karachi, Davao, Manila, Bangkok, Hanoi, and Ho Chi Minh City. \textbf{From Latin America (and the Caribbean):} Buenos Aires, La Paz, Belo Horizonte, Brasilia, Rio de Janeiro, Salvador, Sao Paulo, Bogota, Cali, Medellin, San Jose, Guayaquil, Quito, San Salvador, Guatemala City, Tegucigalpa, Acapulco, Cancun, Guadalajara, Leon, Merida, Mexico City, Monterrey, Puebla, Queretaro, Managua, Asuncion, Lima, Santo Domingo, and Port-au-Prince.} \label{Map}
\end{figure*}
}

\subsection*{Data used to model cities}

To quantify the links between urban form and water, we utilise a wide variety of datasets, including satellite-derived, surveys and other geographical datasets described below. This approach has limitations, including biases inherent to global datasets \cite{sartirano2023strengths}, especially in regions where reliable administrative statistics are lacking and often serve as imperfect proxies for urban realities. Yet, these datasets are among the best proxies that can be utilised to compare many cities from many countries.

{
$\circ$ The \textbf{Buildings dataset} is obtained from the Google Open Buildings dataset \cite{sirko2021continental}. It detects buildings using satellite images. We aim to replicate the definition of an urban area applied in the Africapolis dataset \cite{Africapolis}. For Africapolis, the proximity of infrastructure is a criterion used to determine whether a building is part of a city's delineation. For each city, we first delineate its perimeter using detailed information based on the location and footprints of buildings. This definition enables us to apply the same criteria across cities, utilising the proximity of infrastructure so that we can compare the perimeters obtained for cities in Africa, Asia, and Latin America. Other delineations could also be considered, for example, by considering commuters. However, that data is not available for many cities, so it cannot be widely applied in our study.
}

{
$\circ$ The \textbf{Relative Wealth Index} database is a publicly available dataset that provides local-level estimates of wealth for all low- and middle-income countries \cite{chi2022microestimates}. The Relative Wealth Index measures the asset-based wealth of approximately 135 countries worldwide at a resolution of 2.4 km. The estimates are based on household data obtained by the DHS program, as well as other sources such as mobile phone and social media connectivity data \cite{chi2022microestimates}.
}

{
$\circ$ The \textbf{Critical Infrastructure Index} is a harmonised spatial dataset for representing the intensity of infrastructure \cite{nirandjan2022spatially}. It was produced by aggregating high-resolution geospatial data from OpenStreetMap \cite{OpenStreetMap}. The data for critical infrastructure was obtained from 39 layers categorised into seven overarching systems: Energy, Water Supply, Solid Waste, Transportation, Telecommunications, Healthcare, and Education. The index is expressed as a dimensionless value, ranging from 0 (no intensity) to 1 (highest intensity). Each layer is listed below.

\begin{itemize}
\item \textbf{Energy} - cable, line, minor line, plant, substation, power tower, power pole. 
\item \textbf{Water supply} - water tower, water well, reservoir covered, water works, reservoir. 
\item \textbf{Solid waste} - landfill, waste transfer station, water waste treatment plant. 
\item \textbf{Transportation} - railway, primary, secondary, tertiary, airport.
\item \textbf{Telecommunications} - communication tower, mast.
\item \textbf{Healthcare} - clinic, doctors, hospital, dentist, pharmacy, physiotherapist, alternative, laboratory, optometrist, rehabilitation, blood donation, birthing centre. 
\item \textbf{Education} - college, kindergarten, library, school, university. 
\end{itemize}
}

{
There are many issues with crowdsourced geodata, including biases in reporting \cite{prieto2022constructing}. The data typically has better coverage in regions with a higher number of internet users, so it is possible that the data does not capture all parts of the critical infrastructure. However, the Critical Infrastructure Index captures, at a more general level, how close any location is to various types of aggregate infrastructure. Additionally, the data has a global coverage, so it is possible to obtain an index for all locations within a city.
}

{
$\circ$ The \textbf{Joint Monitoring Programme for Water Supply, Sanitation and Hygiene (JMP)} dataset provides an estimate of the population with access to piped water \cite{JMPDatabase}. The data is produced by a joint effort of the World Health Organization, UNICEF, UNESCO and others. For each city, access to piped water is assumed to be the share of the population in urban areas of the country with access to this service.
}

{
$\circ$ The \textbf{Demographic and Health Surveys Program (DHS)} is a survey conducted at the household level in low and middle-income countries \cite{DHSData}. The project is funded by the United States Agency for International Development (USAID) with contributions from other donors such as UNICEF, UNFPA, WHO, and UNAIDS. For each household, the collected information indicates whether it has access to water and sanitation services and the location (GPS coordinates) of its neighbourhood. Using the remoteness of each household, we estimate the probability of having access to water and sanitation services.
}

{
$\circ$ The \textbf{IBNet Tariffs database} (International Benchmarking Network for Water and Sanitation Utilities) is produced by the World Bank and aggregates water tariffs \cite{WaterTariffs}. The database contains water and wastewater tariffs at the city level in a comparable US\$ format per city. The reported numbers are only related to tariffs, not to affordability, so the numbers must be interpreted with care. 
}

{
$\circ$ The \textbf{Gridded Gross Domestic Product dataset} provides an extensive dataset of GDP per person at the admin-2 level \cite{kummu2025downscaled}. To spatially downscale the data, it is assumed to be uniform within each area, and then, a combination of extrapolation techniques and machine learning algorithms was applied \cite{kummu2025downscaled}. The resulting dataset is provided as a raster image, enabling spatial analysis of estimated GDP per person across locations. Each pixel reflects the GDP per capita of its corresponding admin-2 unit. Most metropolitan areas in our analysis span multiple admin-2 regions, each with its GDP per person estimate (Supplementary Figure \ref{DifferentValues}). Therefore, we can analyse several distinct income values within a single urban area. For example, Bangkok comprises 49 admin-2 units, whilst Mexico City consists of 37. To examine how income varies with remoteness, we analysed the GDP per person at the pixel level, where the distance from the urban core can be estimated. For each location, we extracted the associated GDP per capita value, which often appears multiple times, especially in large admin-2 areas that cover many pixels.

\begin{figure}[h!] \centering
\includegraphics[width = 0.5\linewidth]{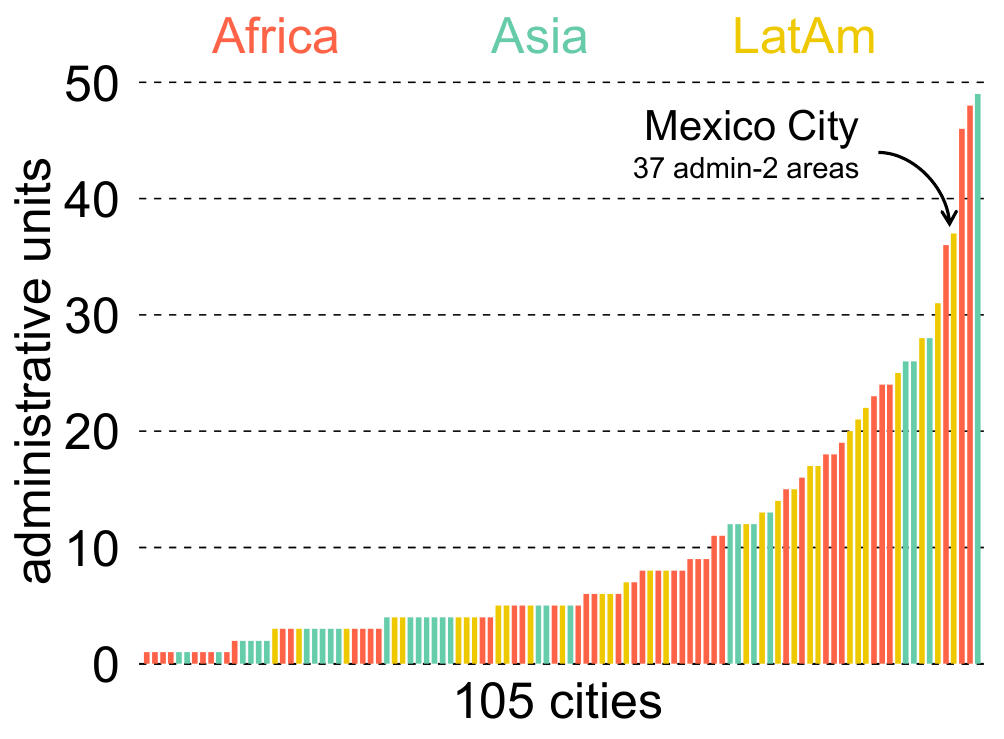}
\caption{Number of unique administrative units (admin-2 level) per city, colored by continent. Each vertical bar represents one city, and the height of the bar corresponds to the number of distinct admin-2 units. Of the cities analysed, 90\% span more than a single admin-2 area, so it is possible to get an approximate idea of how the GDP per person changes across distinct locations of a city.} \label{DifferentValues}
\end{figure}

}

{
$\circ$ The \textbf{Gridded Population of the World (GPW)} models the distribution of human population (counts and densities) for all locations at the most detailed spatial resolution available from the results of the 2010 round of Population and Housing Censuses, which occurred between 2005 and 2014. The input data are extrapolated to produce population estimates for different years \cite{PopulationData, niva2023world}.
}

{
Some of the datasets are not available for all cities in the study (Supplementary Table \ref{CityLevelTable}). In those cases, we dropped the cities from the specific part of the study and reported the results for those for which the data is available. Other datasets, such as the gridded population dataset and GDP, are available across all cities.

\begin{table}
    \centering
    \begin{tabular}{rc|cccc}
    Data &   source  & cities & Africa & Asia & LatAm\\
        \hline
        Buildings &  \cite{sirko2021continental} & 105 & 45 & 30 & 30 \\ 
        Gridded GDP per person & \cite{kummu2025downscaled} & 105 & 45 & 30 & 30 \\
        Relative Wealth Index & \cite{chi2022microestimates} & 105 & 45 & 30 & 30 \\
        Water Tariffs & \cite{WaterTariffs}      & 85 & 36 & 22 & 27 \\
        DHS &  \cite{DHSData}                    & 75 &  40 & 25 & 10 \\
         \hline 
             \end{tabular}
    \caption{Number of cities for each dataset in the data is available.}
    \label{CityLevelTable}
\end{table}
}

{
Although the data is available only for some cities at the city level, the number of observations that compose each dataset is thousands or millions (Supplementary Table \ref{OtherLevelTable}). For example, we rely on the footprint of more than 183 million buildings to delineate the polygon of each city and to determine the constructed surface within walking distance \cite{sirko2021continental}. 

\begin{table}
    \centering
    \begin{tabular}{r|cccc}
    Data  & N & Africa & Asia & LatAm\\         
         \hline
         Buildings (millions)& 183.4 & 61.2 & 60.4 & 61.8\\
         Relative Wealth Index (thousands) & 19.7 & 7.3 & 6.6 & 5.8 \\
         DHS (thousands) & 180 & 85 & 62 & 33 \\         
    \end{tabular}
    \caption{Number of observations used for each table}
    \label{OtherLevelTable}
\end{table}
}

\subsection*{Water scarcity and sponge cities}

{
Most urban water is piped from reservoirs and water basins rather than sourced directly from rainfall or soil moisture. However, the conditions related to water in cities today are vastly heterogeneous. It is possible to look at water stress (as Figure 1 in the manuscript, with data from \cite{wri_aqueduct_2023}). However, other elements may be considered. In terms of the terrain, for example, it is possible to determine the soil's capacity to retain water (Supplementary Figure \ref{AridityRain}). The soil's capacity to retain water plays a vital role in urban areas, not only as an environmental indicator but as a cornerstone for managing stormwater runoff, reducing flood risks, and improving the availability and reliability of water for urban services such as sanitation and domestic supply \cite{shi2016disturbed}. In rural or uninhabited regions, the soil's capacity to retain water is mainly used to assess vegetation dynamics and ecosystem functions. However, in cities, it highlights the potential to support sustainable water management in densely populated and infrastructure-intensive settings. The soil's capacity to retain water assesses evapotranspiration processes and rainfall deficit for potential vegetative growth and can be determined with the Global Aridity Index \cite{zomer2022version}.

\begin{figure}[ht!] \centering
\includegraphics[width = 0.95\linewidth]{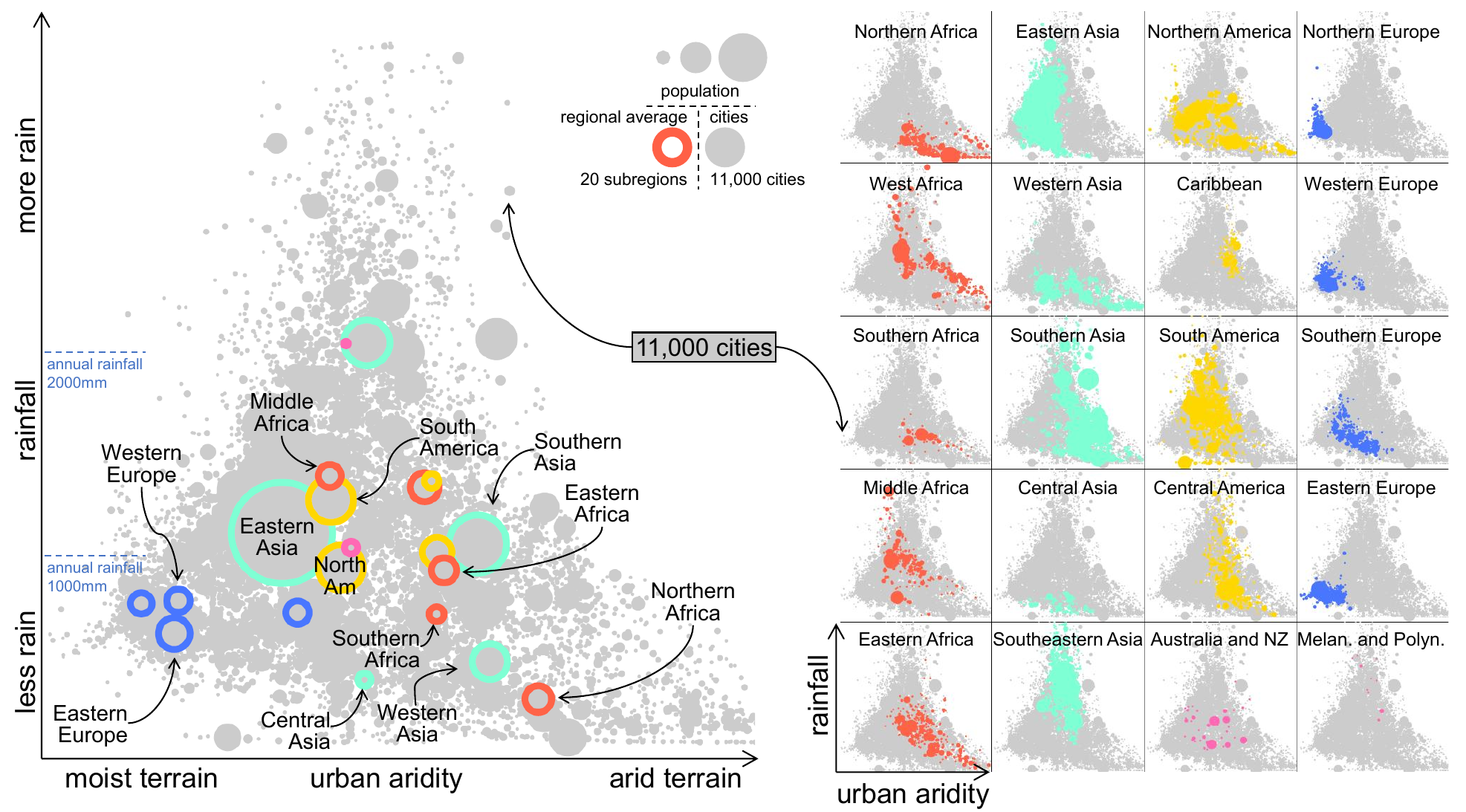}
\caption{We estimate urban aridity by detecting the soil capacity to retain water (horizontal axis) and compare it with the yearly precipitation (vertical axis). The rainfall is the average monthly precipitation between 1970 and 2000, aggregated to a yearly value \cite{fick2017worldclim}. Right - We compute both values for 11,000 cities (each represented by a different disc). The colour corresponds to the region, and the size of the disc corresponds to the population. Left - The rainfall and aridity of each region are the average of cities weighted by the population.  } \label{SMAridityRain}
\end{figure}
}

{
There are vast differences in terms of the rain and the terrain among cities. In most European cities, for example, the terrain is favourable for water retention, so even with low rainfall, the terrain absorbs the moisture (Supplementary Figure \ref{SMAridityRain}). In South America, there is more aridity, but it is also rainier, resulting in better conditions for ensuring access to water. However, there are cities, mainly in Africa, Asia, and parts of America, where rain is not very frequent, but when it does rain, the terrain does not retain the water. In those cities, granting access to water is more challenging.
}

{
There is increasing evidence showing that enhancing the capacity to retain water in urban soils can directly contribute to greater water availability, improved resilience to climate impacts, and more equitable access to water and sanitation services, particularly in marginalised communities \cite{cotler2024ecosystem}. China has led the way in implementing the sponge-city concept \cite{guo2022urbanization}. This approach aims to boost soil and surface water retention through interventions such as permeable pavements, green roofs, and urban wetlands \cite{guo2022urbanization}. These measures reduce surface runoff, recharge groundwater, and lower the costs of water extraction and distribution for urban utilities and state-owned enterprises, helping ensure that water resources are available across both time and space. Several German cities are also adopting sponge-city principles to address urban flooding and improve water quality. Through the use of porous surfaces and green infrastructure, they are expanding service delivery and creating more adaptive, sustainable urban environments \cite{shi2016disturbed}. These strategies are particularly relevant for rapidly urbanising regions in Africa and India, where cities often expand into areas with low aquifer permeability. In such contexts, improving the soil’s capacity to retain water could significantly reduce water loss, lower flood risk, and help bridge the gap in water and sanitation services. Enhancing urban soil water retention helps to mitigate water stress and foster more equitable, resilient cities \cite{cotler2024ecosystem}.
}

\subsection*{Distribution of building size across cities}

{
The infrastructure in some cities is primarily composed of small buildings. For all cities, we consider the footprint of all their buildings and analyse their distribution. Particularly, buildings have a footprint of only a few m$^2$ across some African cities (Supplementary Figure \ref{BuildingComposition}).

\begin{figure}[h!] \centering
\includegraphics[width = 0.65\linewidth]{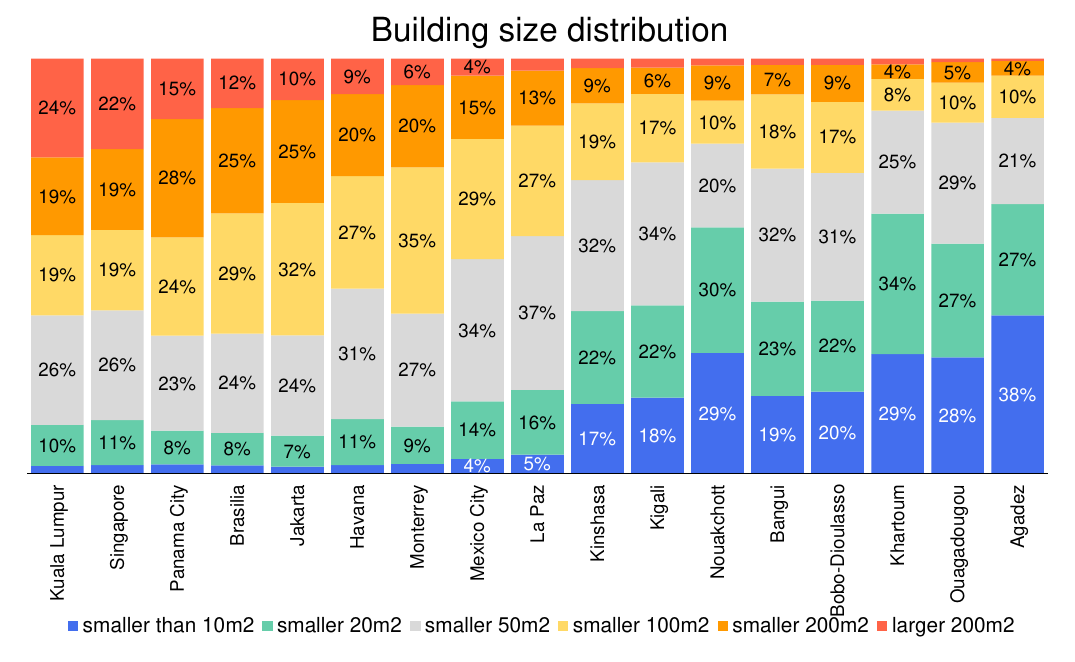}
\caption{Distribution of building size across selected cities. Poor cities are characterised by an abundance of small buildings.} \label{BuildingComposition}
\end{figure}
}

\subsection*{Walking distance and infrastructure}

{
There is no formal definition of a reasonable walking distance. Additionally, the distance that a person walks depends on many aspects, including the age and the health condition of the person, and even the purpose of the trip. For example, a travel survey found that the median distance walked to or from a public transit stop was 530 m, but to shops was 770 m \cite{SUGIYAMA2019100621}. Thus, we take $\tau = 500$ m as a reasonable walking distance. However, we also adjust this threshold to determine its relevance. 
}

{
For each pixel, we quantified the number of buildings, the average area of buildings and the constructed surface of buildings within walking distance. Here, we demonstrate that the distance threshold ($\tau$) has minimal impact. For example, if instead of taking $\tau =500$ m as the walking distance, we take $\tau =400$ m or $\tau =600$ m, the number of buildings and the constructed surface vary (Supplementary Figure \ref{WalkingDistance}). 

\begin{figure}[h!] \centering
\includegraphics[width = 0.65\linewidth]{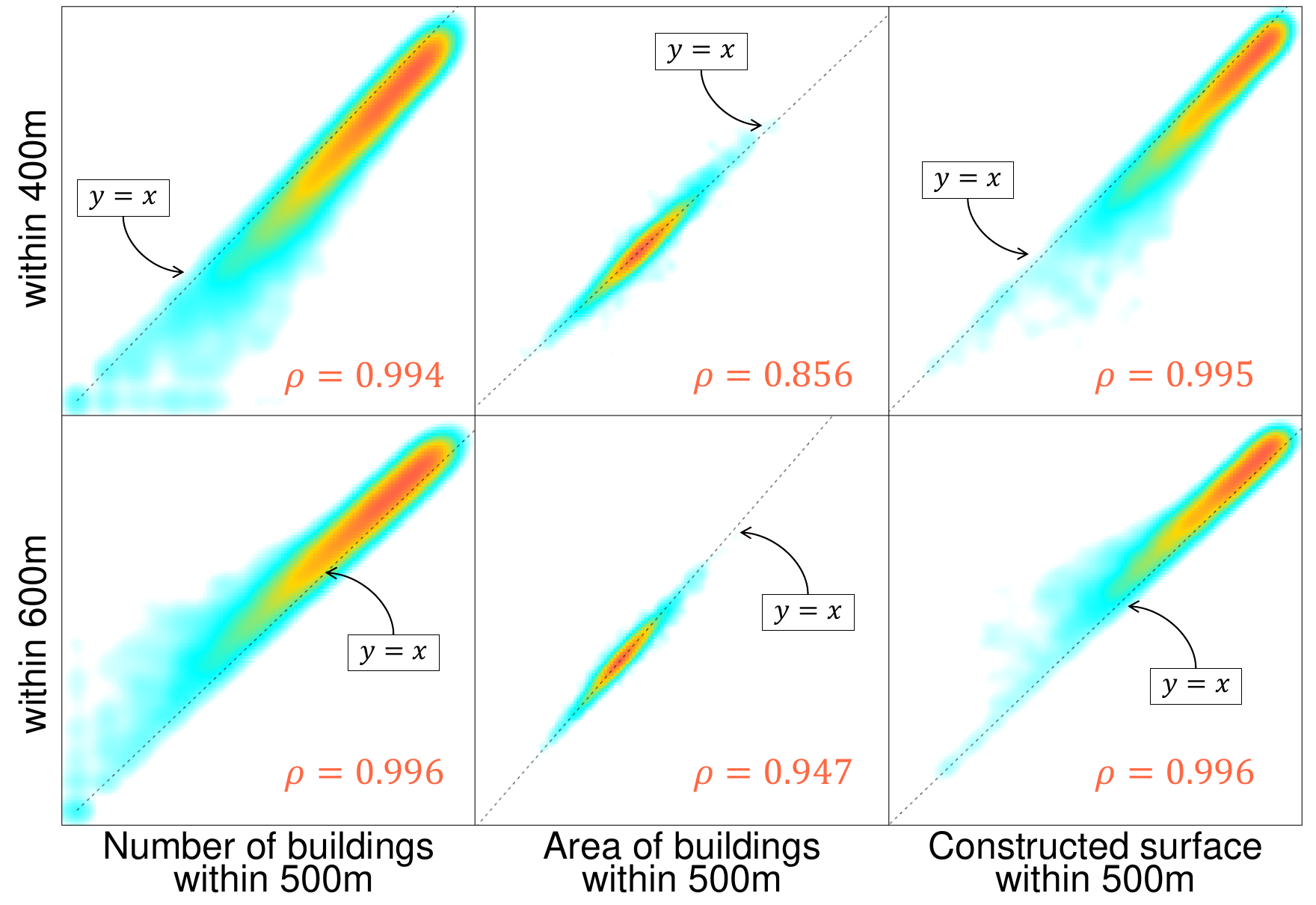}
\caption{Number of buildings (left), the mean area of buildings (centre) and constructed surface (right) considering $\tau =500$ m for walking distance (horizontal axis) $\tau =400$ m (top vertical axis) or $\tau =600$ m (bottom vertical axis).} \label{WalkingDistance}
\end{figure}
}

{
Although the number of buildings within $\tau =400$ m is smaller than the number of buildings within $\tau =500$ m, the correlation between the two variables is 0.994 (Supplementary Figure \ref{WalkingDistance}). We obtain the same results for longer walking distances and other metrics, such as the average size of buildings and the constructed surface. Therefore, the distance threshold for what is considered a reasonable walking distance has a limited impact on how a pixel is classified and its implications in terms of remoteness or water infrastructure.
}

\subsection*{Identifying the centre of a city}

{
Cities may be highly heterogeneous in terms of accessibility, commuting distances, housing prices, pollution, and more. A technique to analyse that heterogeneity is based on the distance to the centre \cite{lemoy2020evidence}. Although cities may be polycentric (for example, with new business districts in remote locations), those centres do not accurately capture how the city expanded historically from its centre or how some neighbourhoods close to it have better provisions. 
}

{
Selecting the centre of a city is not always obvious. We manually identify the centre as the location of the main square, church, mosque, or central transport station. In Luanda, for example, we have identified the National Assembly of Angola as the city centre. Other reasonable options include the Banco Nacional de Angola and Estação Central de Luanda (Supplementary Figure \ref{DifferentCentres}). The distances from Estação de Viana (a public transport station and candidate for a new international airport) are 18.9 km, 18.7 km, and 18.3km. Thus, Viana is a remote location regardless of the specific choice of the city centre. We test the impact of picking a distinct centre by delineating the central part of the city, defined by the circumference corresponding to the ring with $R_{i,j} \leq 3$ (Supplementary Figure \ref{DifferentCentres}). 

\begin{figure}[h!] \centering
\includegraphics[width = 0.75\linewidth]{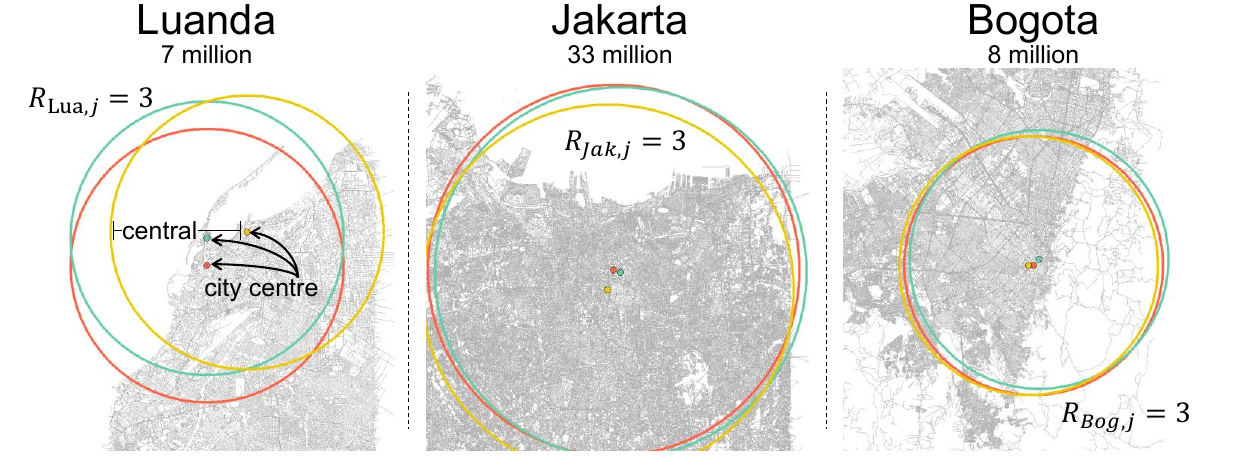}
\caption{Impact of selecting different locations as the city centre in Luanda, Jakarta and Bogota. The coloured circles correspond to the parts of the city considered central (with $R_{i,j} \leq 3$) according to those options. } \label{DifferentCentres}
\end{figure}
}

{
In Jakarta, we picked the Monas National Monument (and home to the Museum Sejarah Nasional), but other candidates include the National Museum of Indonesia, the National Gallery of Indonesia or Plaza Indonesia. The distances to the Soekarno-Hatta International Airport are 19.4 km, 19.9 km, 20.4 km and 20.1 km. In Bogota, for example, three options are available: Plaza Bolívar, Museo del Oro (located 500 m north), or Parque Metropolitano (located 600 m west). 
}

{
We evaluate the criticality of the definition of city centre by quantifying the overlap between different central areas (the extent to which the rings with $R_{i,j} \leq 3$ intersect). A high degree of overlap indicates that, regardless of the definition used, the delineation of central, inter-urban, distant, or peri-urban areas remains broadly consistent. In Luanda, the overlap of areas identified as the city centre ranges from 79 to 92\%, depending on the definition applied. In Jakarta, this overlap ranges from 92 to 97\%, and in Bogota, from 93 to 98\%.
}

{
Similarly, it has been shown that when the centre of a city is identified as the city hall and then arbitrarily changed to other locations, the impact on a radial analysis of the city is minor \cite{lemoy2020evidence}. Thus, although the precise definition of the city centre is often ambiguous, it has a minor impact on the parts of the city that are identified as remote or periurban. 
}

\subsection*{Remoteness within a city}

{
Consider the location $j$ in the city $i$. Let $D_{i,j}$ be the distance in km to the centre of that city. We define the \emph{remoteness} of location $j$ as
\begin{equation} 
R_{i,j} = \frac{1000 D_{i,j}}{\sqrt{P_i}},
\end{equation}
where $P_i$ is the population of city $i$. The remoteness captures how a city expands, usually around the centre. The principle is rooted in geographical economic theory, which considers how certain elements (such as price and demand for real estate) change as the distance from the centre increases \cite{alonso1964location}. This process of expansion has been characterised into different stages of compaction and dispersion based on different spatial regimes and population growth \cite{land9060200, seto2010new}. The idea is to characterise locations that are at a similar distance from the centre of each city. However, since distance in cities varies considerably depending on their size, we construct remoteness to discount any direct impacts of population size on distance metrics. The principle behind the remoteness is to construct a metric for distance that can be compared between cities of different sizes. If each person occupies the same surface of a city (say 10 m$^2$), then the denominator is proportional to the square root of the area of that city, (so what could be called the ``radius'' of the city) \cite{lemoy2021radial}. Since remoteness is the result of dividing a distance (in km) by the square root of an area (which is also in distance units), the result is an index that has no units because they cancel out when calculating the value. We use the remoteness $R_{i,j}$ of location $j$ in city $i$ as the critical variable to quantify the relative distance across cities. 
}

{
For most cities, remoteness varies from 0 at the centre to some maximum value that changes considerably. For example, in Dhaka, the maximum remoteness is 5, similar to Ahmedabad and Bangalore. However, in other cities, such as Kisumu, Acapulco, and Kigali, some locations have a remoteness of over 30. To consider an objective partition of different values of remoteness, we use a classification tree, where we group different values of the constructed surface within walking distance ($\phi$) and use remoteness as the variable that divides it \cite{breiman1984classification}. The principle here is to detect different intensities of urbanisation (so, how much is constructed within walking distance) as we move away from the city centre. The objective of the classification is to group the distinct values of remoteness, characterising each group by its various levels of infrastructure. We use a classification tree to divide the remoteness of a city using the constructed surface within walking distance as the element which varies depending on remoteness \cite{CiteRpart, RCoreTeam2018}.

To explain the principle behind the classification, consider the following 12 hypothetical observations or pixels (Supplementary Table \ref{SimPixels}). For each observation, we take the remoteness $R$ and the constructed surface within walking distance $S$ in \%. The first observation, for example, has a remoteness of 0.8 and 45.2\% of the surface within walking distance is constructed. The objective of a classification tree is to group the observations that have a more similar constructed surface within each group (initially into two groups) based on the values of $R$. That is, we aim to find a cut-off point $R_0$ such that two groups are formed, one with $R \leq R_0$ and one with $R > R_0$. The criterion to select a threshold is to minimise the variance of \( S \) within each group.  

\begin{table}
\begin{center}
\begin{tabular}{c|cccccccccccc}
\hline
\textbf{Observation} & 1 & 2 & 3 & 4 & 5 & 6 & 7 & 8 & 9 & 10 & 11 & 12 \\
\hline
Remoteness $R$ & 0.8 & 1.5 & 2.4 & 3.5 & 4.2 & 5.8 & 6.7 & 7.5 & 8.8 & 9.4 & 10.2 & 11.5 \\
Const. surface $\phi$ & 45.2 & 42.8 & 40.1 & 28.0 & 25.7 & 21.3 & 18.9 & 16.5 & 14.2 & 9.5 & 7.1 & 5.3 \\
group & ---- & ---- & ---- & . & . & . & . & . & . & . & . & .\\
\hline
\end{tabular}
\caption{Simulated remoteness $R$ and constructed surface $\phi$ for 12 pixels.}
\label{SimPixels}
\end{center}
\end{table}

In the example above, we can compute the combined variance of the two groups obtained by dividing the 12 observations using a different cut-off point $R_0$. The combined variance is smallest when we take \( R_0 = 3 \), separating observations with the highest constructed surface from the rest. Choosing \( R_0 = 3 \) as the threshold isolates the three highest values of the constructed surface \( \phi \), which represent dense urban areas. This results in a low within-group variance for both resulting groups. If instead we used \( R = 4 \), for example, the lower group would include observation 4, a mid-range value (of $\phi = 28.0$), increasing the variance and making the classification less effective. This illustrates how the tree selects split points that best separate different urban forms based on their constructed surface.
}

{
In our analysis, we followed a similar procedure but with nearly 20,000 pixels, using the remoteness $R$ to classify the constructed surface within walking distance $\phi$. For all pixels observed, we classify them so that each group can be considered homogeneous and different from the observations of other groups. The result of the classification tree can be interpreted as follows. On average, among the urban areas considered, 14\% of the surface within walking distance is constructed. If we want to consider only two intensities of constructed surface, a cut-off point of $R_0 = 5$ divides cities into two groups. Pixels with $R_{i,j} \leq 5$ have 23\% constructed surface within walking distance, but pixels with $R_{i,j} > 5$ have only 9.6\% (Supplementary Figure \ref{ClassTree}). If we further subdivide into groups, we obtain four groups with the corresponding cut-off points.

\begin{figure}[h!] \centering
\includegraphics[width = 0.75\linewidth]{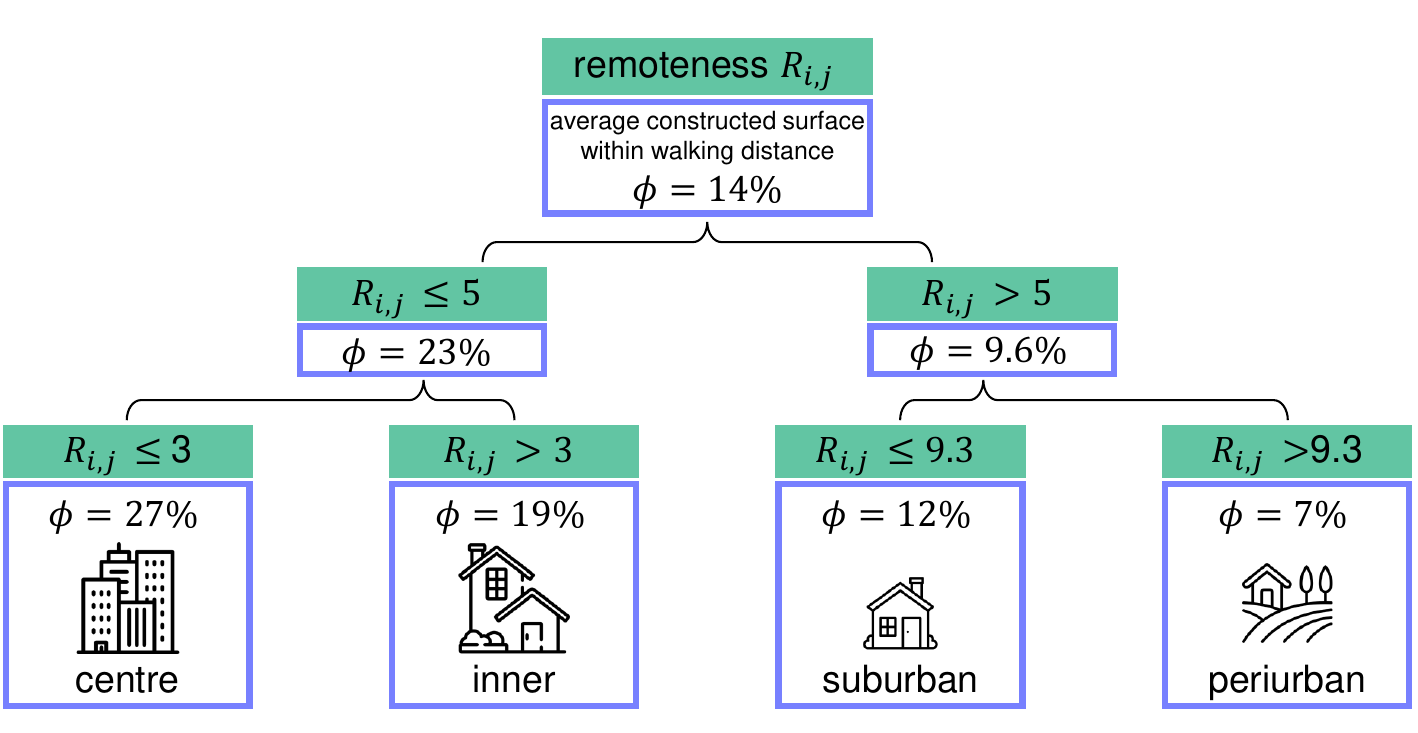}
\caption{Classification tree of the constructed surface within walking distance, $\phi$ based on the remoteness $R_{i,j}$ of each location.  } \label{ClassTree}
\end{figure}

The classification tree divides the remoteness into four parts:
\begin{itemize}
\item $R_{i,j} \leq 3$ - \textbf{central}. 
\item $R_{i,j} \in (3,5]$ - \textbf{inner}. 
\item $R_{i,j} \in (5,9.3]$ - \textbf{suburban}. 
\item $R_{i,j} > 9.3$ - \textbf{peri-urban}. 
\end{itemize}
}

{
In the central area ($R_{i,j} \leq 3$), the average constructed surface within walking distance is $\phi = 27\%$, but it drops to 19\% for the intermediate areas, decreases further to 12\% for the distant areas and then, if $R_{i,j} > 9.3$, then it drops to 6.9\%. With this classification, we divide each city into a central area and three consecutive rings.
}

\subsection*{Increasing demand for water per person}

{
In 1960, less than 10\% of the world population lived in areas with chronic water scarcity, but that number has rapidly increased \cite{kummu2010physical}. In most parts of the world, water scarcity shows an increasing trend \cite{kummu2010physical}. It directly affects people, leaving many without access to an improved water source that is accessible on premises, available when needed, and free from faecal and priority chemical contamination (safely managed). Even in wealthy countries like the US, lack of access in some areas is a concern \cite{meehan2020geographies}. Additionally, there are many heterogeneities in what people demand globally. For example, in Mexico City, daily water consumption ranges from less than 10 litres in remote, poor neighbourhoods to 650 litres per person in central, wealthy neighbourhoods \cite{medina2022spatial}.
}

{
Some global trends help us understand the collective volume of what we consume. Although most water used for cattle and some water used for industry is not part of the demand for urban areas, it still puts pressure on systems used to satisfy urban demands. In the US, for example, 39.6\% of the water withdrawal is for agriculture, 47.2\% is industrial, and only 13.2\% is municipal water withdrawals \cite{aquastat2025aquastat}. However, in Mexico, 75.7\% of the water withdrawals are for agriculture, 9.5\% are for industry and 14.7\% are municipal water withdrawals \cite{aquastat2025aquastat}. 
}

{
There are also some relevant trends in the way we demand resources globally. In 1960, a person consumed roughly 23 kg of meat each year (including poultry, beef, pork, and others) \cite{owid-meat-production}. By 2022, a person consumed, on average, more than 44kg. However, at the same time, the world's population increased from 3 billion to 8 billion, resulting in a fivefold increase in global meat demand during that period. Additionally, meat consumption tends to be much higher in wealthier countries and increases as countries become richer \cite{owid-meat-production}. The production of meat is one of the largest pollutants and also one of the most water-demanding activities worldwide. Besides meat, other water-intensive products are increasing the per-person demand. For example, the global milk production has increased faster than the world's population.  
}

{
Besides water for food consumption, other products with increasing demand are also water-intensive. For example, manufacturing a car uses roughly 150 m\textsuperscript{3} of water \cite{WaterCars}. In 1960, the world manufactured approximately 16 million vehicles, but today, we manufacture six times more cars each year. Even if a person drinks only a small amount of water a day, they are likely to increase their water consumption in the coming years. Thus, more people around the world could demand more water, even if that demand is not in their own city.  
}

\subsection*{Linear models at the city level}

Here, we present the coefficients of a sequence of multivariate linear models at the city level. 

\textbf{Model the availability of Piped water} 

{
We model the availability of piped water at the city level using sparseness, population, and the continent of the city, with Africa as the baseline (Supplementary Table \ref{Tariff.models}). Results show that cities with more sparseness have a smaller number of people with piped water. We also include the population and climatic conditions of the city. For the model with more variables considered, results show that cities with more critical infrastructure provide water for a higher number of people.

\renewcommand{\arraystretch}{0.8}
{\scriptsize
\begin{table}[h]
\begin{center}
\begin{tabular}{l c c c}
 & Model 1 & Model 2 & Model 3 \\
\hline
(Intercept)             & $16.589^{***}$ & $-0.890$      & $0.017$       \\
                        & $(0.435)$      & $(0.912)$     & $(1.241)$     \\
log(sparseness)     & $-1.212^{***}$ &     $-0.046$      & $0.100$       \\
                        & $(0.245)$      & $(0.138)$     & $(0.142)$     \\
log(population)         &                & $1.027^{***}$ & $0.973^{***}$ \\
                        &                & $(0.053)$     & $(0.059)$     \\
continent - Asia           &                & $-0.143$      & $0.007$       \\
                        &                & $(0.139)$     & $(0.176)$     \\
continent - Latin America  &                & $0.482^{***}$ & $0.500^{**}$  \\
                        &                & $(0.122)$     & $(0.150)$     \\
MeanCritInf             &                &               & $2.986^{**}$  \\
                        &                &               & $(1.007)$     \\
FootpLargeBuildings050 m &                &               & $-1.465$      \\
                        &                &               & $(0.744)$     \\
aridity                    &                &               & $0.000$       \\
                        &                &               & $(0.000)$     \\
rain                    &                &               & $0.006$       \\
                        &                &               & $(0.004)$     \\
dryness                     &                &               & $-0.000$      \\
                        &                &               & $(0.000)$     \\
\hline
R$^2$                   & $0.192$        & $0.840$       & $0.866$       \\
Adj. R$^2$              & $0.184$        & $0.833$       & $0.852$       \\
Num. obs.               & $105$          & $105$         & $99$          \\
\end{tabular}
\caption{Models for piped water}
\label{PipedWater}
\end{center}
\end{table}
}
\renewcommand{\arraystretch}{1}
}

\clearpage

\textbf{Model the Water tariffs} 

{
We model water tariffs in cities by considering the sparseness, city size and the continent, added sequentially (Supplementary Table \ref{Tariff.models}). Results show that keeping everything else constant, sparseness is correlated with higher water tariffs. Additionally, in cities with larger populations, tariffs tend to be smaller. As has been observed, sprawled, elongated cities have lower resilience and sustainability \cite{li2020influence}.

{
\renewcommand{\arraystretch}{0.8}
{\scriptsize
\begin{table}[h]
\begin{center}
\begin{tabular}{l c c c}
 & Model 1 & Model 2 & Model 3 \\
\hline
(Intercept)            & $0.614$       & $2.497$      & $1.929$        \\
                       & $(0.346)$     & $(1.668)$    & $(1.506)$      \\
log(sparseness)    & $0.815^{***}$ & $0.661^{**}$ & $0.334$        \\
                       & $(0.196)$     & $(0.237)$    & $(0.226)$      \\
log(population)        &               & $-0.107$     & $-0.024$       \\
                       &               & $(0.093)$    & $(0.086)$      \\
continent - Asia          &               &              & $-0.851^{***}$ \\
                       &               &              & $(0.241)$      \\
continent - Latin America &               &              & $0.222$        \\
                       &               &              & $(0.186)$      \\
\hline
R$^2$                  & $0.186$       & $0.200$      & $0.370$        \\
Adj. R$^2$             & $0.175$       & $0.179$      & $0.336$        \\
Num. obs.              & $78$          & $78$         & $78$           \\
\end{tabular}
\caption{Models for Water Tariffs}
\label{Tariff.models}
\end{center}
\end{table}
}
\renewcommand{\arraystretch}{1}
}
}

\clearpage

\textbf{Model the average Critical Infrastructure} 

{
We model the mean Critical Infrastructure in a city using sparseness, population, and the continent of the city, taking Africa as the baseline. We add variables in a sequence (Supplementary Table \ref{CriticalInf}). Results show that with more sparseness, there is less critical infrastructure. Also, cities with larger populations tend to have more critical infrastructure, and finally, cities in both Asia and Latin America have more average infrastructure. 

{
\renewcommand{\arraystretch}{0.8}
{\scriptsize
\begin{table}[h]
\begin{center}
\begin{tabular}{l c c c}
 & Model 1 & Model 2 & Model 3 \\
\hline
(Intercept)            & $-0.947^{***}$ & $-4.091^{***}$ & $-4.103^{***}$ \\
                       & $(0.196)$      & $(0.846)$      & $(0.777)$      \\
log(sparseness)    & $-0.682^{***}$ & $-0.451^{***}$ & $-0.436^{***}$ \\
                       & $(0.111)$      & $(0.120)$      & $(0.117)$      \\
log(population)        &                & $0.184^{***}$  & $0.169^{***}$  \\
                       &                & $(0.048)$      & $(0.045)$      \\
continent - Asia          &                &                & $0.240^{*}$    \\
                       &                &                & $(0.119)$      \\
continent - Latin America &                &                & $0.469^{***}$  \\
                       &                &                & $(0.106)$      \\
\hline
R$^2$                  & $0.281$        & $0.376$        & $0.484$        \\
Adj. R$^2$             & $0.274$        & $0.363$        & $0.462$        \\
Num. obs.              & $99$           & $99$           & $99$           \\
\end{tabular}
\caption{Models for Critical Infrastructure}
\label{CriticalInf}
\end{center}
\end{table}
}
\renewcommand{\arraystretch}{1}
}
}

\clearpage

\textbf{Model the sparseness}

{
Finally, at the city level, we model the sparseness. We take the logarithmic transformation of the sparseness at the city level and include, first, the population, then the continent, and finally, the number of buildings (Supplementary Table \ref{Sparseness.models}). Results show that larger cities tend to have less sparseness, and Asian cities are also less sparse. When the number of buildings is included in the model, the results indicate that cities tend to be sparser with more buildings. However, the coefficient for population is compensated for in large cities. Roughly, for every four people in a city, there is one additional building \cite{prieto2023scaling}.

\renewcommand{\arraystretch}{0.8}
{\scriptsize
\begin{table}[h]
\begin{center}
\begin{tabular}{l c c c}
 & Model 1 & Model 2 & Model 3 \\
\hline
(Intercept)            & $4.797^{***}$  & $4.139^{***}$  & $3.091^{***}$  \\
                       & $(0.506)$      & $(0.510)$      & $(0.556)$      \\
log(population)        & $-0.206^{***}$ & $-0.156^{***}$ & $-0.341^{***}$ \\
                       & $(0.034)$      & $(0.035)$      & $(0.059)$      \\
continent - Asia          &                & $-0.329^{***}$ & $-0.278^{**}$  \\
                       &                & $(0.095)$      & $(0.090)$      \\
continent - Latin America &                & $-0.002$       & $-0.074$       \\
                       &                & $(0.088)$      & $(0.085)$      \\
log(NBuildings)        &                &                & $0.275^{***}$  \\
                       &                &                & $(0.073)$      \\
\hline
R$^2$                  & $0.264$        & $0.354$        & $0.433$        \\
Adj. R$^2$             & $0.257$        & $0.335$        & $0.410$        \\
Num. obs.              & $105$          & $105$          & $105$          \\
\end{tabular}
\caption{Models for Remoteness}
\label{Sparseness.models}
\end{center}
\end{table}
}
\renewcommand{\arraystretch}{1}
}

\clearpage

\subsection*{Models at the pixel level}

Here, we present the coefficients of a sequence of linear models at the pixel level. 

\textbf{Model the GDP per person at the micro level} 

{
We construct three regression models to characterise the GDP per person at the pixel level (Supplementary Table \ref{GDPPP}). We include the continent (with Africa as a baseline). We also include the remoteness, and sequentially, we add the population (on a log scale), the number of buildings (scaled by a $1/1000$ factor), the constructed surface within walking distance (as a per cent) and the average area of buildings (scaled by a $1/10,000$ factor). The results of the three models indicate that with increased remoteness, there is a lower GDP per person, and also that locations with a greater constructed surface area within walking distance tend to be wealthier.

\renewcommand{\arraystretch}{0.8}
{\scriptsize
\begin{table}
\begin{center}
\begin{tabular}{l c c c}
\hline
 & Model 1 & Model 2 & Model 3 \\
\hline
(Intercept)            & $9.072^{***}$  & $7.006^{***}$  & $7.008^{***}$  \\
                       & $(0.011)$      & $(0.077)$      & $(0.077)$      \\
continent - Asia          & $0.716^{***}$  & $0.600^{***}$  & $0.523^{***}$  \\
                       & $(0.012)$      & $(0.012)$      & $(0.013)$      \\
continent - Latin America & $1.037^{***}$  & $1.023^{***}$  & $0.956^{***}$  \\
                       & $(0.012)$      & $(0.011)$      & $(0.012)$      \\
$R_{i,j}$           & $-0.031^{***}$ & $-0.026^{***}$ & $-0.023^{***}$ \\
                       & $(0.001)$      & $(0.001)$      & $(0.001)$      \\
log(population)        &                & $0.132^{***}$  & $0.127^{***}$  \\
                       &                & $(0.005)$      & $(0.005)$      \\
nbuildings walking distance       &                &                & $-0.069^{***}$ \\
                       &                &                & $(0.007)$      \\
surface walking distance          &                &                & $1.159^{***}$  \\
                       &                &                & $(0.068)$      \\
meanArea walking distance         &                &                & $1.755^{***}$  \\
                       &                &                & $(0.275)$      \\
\hline
R$^2$                  & $0.384$        & $0.406$        & $0.419$        \\
Adj. R$^2$             & $0.384$        & $0.406$        & $0.419$        \\
Num. obs.              & $19,677$        & $19,677$        & $19,677$        \\
\hline
\end{tabular}
\caption{Models for GDP PP}
\label{GDPPP}
\end{center}
\end{table}
}
\renewcommand{\arraystretch}{1}
}

\clearpage

\textbf{Model the Relative Wealth Index} 

{
The data cannot be compared directly across countries, as metrics are relative, with a mean value of zero and a standard deviation of one for each country. By including country effects in a model, we capture how the Relative Wealth varies based on remoteness (Supplementary Table \ref{MRWI.models}). Results show that with greater remoteness, there is a lower Relative Wealth Index. Additionally, with less constructed surface area within walking distance, there is a lower Relative Wealth Index. The country effects are not shown (55 countries).

{
\renewcommand{\arraystretch}{0.8}
{\scriptsize
\begin{table}[h]
\begin{center}
\begin{tabular}{l c}
 & Model 1 \\
\hline
(Intercept)                     & $1.441^{***}$  \\
                                & $(0.062)$      \\
country effects                 & ----\\ 
                                & ----\\
$\sqrt{R_{i,j}}$              & $-0.134^{***}$ \\
                                & $(0.003)$      \\
log(population)                 & $-0.024^{***}$ \\
                                & $(0.004)$      \\
surface walking distance                     & $2.03 \times 10^{-6***}$  \\
                                & $(2.82 \times 10^{-8})$      \\
meanArea walking distance                   & $2.34 \times 10^{-5}$       \\
                                & $1.50 \times 10^{-5}$      \\
\hline
R$^2$                           & $0.444$        \\
Adj. R$^2$                      & $0.442$        \\
Num. obs.                       & $19,430$        \\
\end{tabular}
\caption{Model for Relative Wealth Index}
\label{MRWI.models}
\end{center}
\end{table}
}
\renewcommand{\arraystretch}{1}
}

}

\clearpage

\textbf{Model the Critical Infrastructure} 

{
We model the Critical Infrastructure (Supplementary Table \ref{Inf.models}). Results show that the population in larger cities tend to be closer to critical infrastructure. Additionally, with increased remoteness, proximity decreases, and with fewer constructed surfaces within walking distance, there is also less proximity to critical infrastructure.

{
\renewcommand{\arraystretch}{0.8}
{\scriptsize
\begin{table}[h]
\begin{center}
\begin{tabular}{l c c c}
 & Model 1 & Model 2 & Model 3 \\
\hline
(Intercept)            & $-1.166^{***}$ & $0.055^{***}$  & $0.016$        \\
                       & $(0.015)$      & $(0.010)$      & $(0.009)$      \\
log(RelativeDist)      & $-0.682^{***}$ & $-0.064^{***}$ & $-0.050^{***}$ \\
                       & $(0.008)$      & $(0.001)$      & $(0.001)$      \\
continent - Asia          &                & $0.013^{***}$  & $0.006^{***}$  \\
                       &                & $(0.001)$      & $(0.001)$      \\
continent - Latin America &                & $0.039^{***}$  & $0.029^{***}$  \\
                       &                & $(0.001)$      & $(0.001)$      \\
log(population)        &                & $0.011^{***}$  & $0.011^{***}$  \\
                       &                & $(0.001)$      & $(0.001)$      \\
surface walking distance            &                &                & $2.12 \times 10^{-7***}$  \\
                       &                &                & $(5.80 \times 10^{-9})$      \\
meanArea walking distance           &                &                & $1.02 \times 10^{-6}$       \\
                       &                &                & $(3.12 \times 10^{-6})$      \\
\hline
R$^2$                  & $0.297$        & $0.350$        & $0.392$        \\
Adj. R$^2$             & $0.297$        & $0.350$        & $0.392$        \\
Num. obs.              & $19,402$        & $19,402$        & $19,402$        \\
\end{tabular}
\caption{Models for Critical Infrastructure}
\label{Inf.models}
\end{center}
\end{table}
}
\renewcommand{\arraystretch}{1}
}
}

\clearpage

\subsection*{Data availability}

{
The data used for the analysis is available in a Public Repository. It is structured into three separate tables containing information at the city level (105 cities), at the pixel level (20,000 pixels) and the survey level (with nearly 125,000 survey respondents).

https://github.com/rafaelprietocuriel/WaterAndCities

The tables are linked through a unique ID that corresponds to a unique city. The three tables are:

\paragraph{Cities DB}
The table contains the structure for 105 cities in Asia, Africa and Latin America, including variables related to their critical infrastructure, access to water, remoteness, sparseness and water tariffs.

\paragraph{Pixels DB}
The table contains nearly 20,000 observations corresponding to different pixels within a city. All pixels are based on the Relative Wealth Index pixels. They include variables related to rain and critical infrastructure.

\paragraph{Water DB}
The table contains nearly 125,000 survey respondents from the DHS survey. Each observation corresponds to a single survey respondent and records whether the person has access to water and sewage.
}

\section*{Acknowledgements}

RPC is funded by the Austrian Federal Ministry for Climate Action, Environment, Energy, Mobility, Innovation and Technology (2021-0.664.668) and the Austrian Federal Ministry of the Interior (2022-0.392.231).

\section*{Author contributions}

RPC conceived the study, designed the methodology, analysed the results and wrote the manuscript.

PLS compiled the data, analysed the results and wrote the manuscript.

CBV conceived the study, analysed the results and wrote the manuscript.

\section*{Competing interests}

The authors declare that they have no competing interests.

\bibliographystyle{unsrt}

\end{document}